\documentclass[twoside]{article}
\usepackage{astro}
\usepackage[latin1]{inputenc}
\usepackage{a4}
\usepackage{astrobib}
\usepackage{psfig}

\makeatletter

\newcommand{\citeme}[1]{\addtocontents{cit}{\protect\item[{\bf\thesection:}] \protect\citeNP{#1}}}

\renewcommand{\@makecaption}[2]{\small {\bf #1:}~#2}

\begin{document}

\setcounter{page}{15}

\setcounter{figure}{0}
\setcounter{section}{0}
\setcounter{equation}{0}

\pagestyle{myheadings}
\markboth{\hspace*{96mm}{\it \small
Heino Falcke
}\hspace{\fill}}{{\it \small
The\,Silent\,Majority\,--\,Jets\,and\,Radio\,Cores\,from\,Low-Luminosity\,Black\,Holes
}}

\vspace*{-1.4cm}
\thispagestyle{empty}

\noindent
{\scriptsize{\sc Astronomische Gesellschaft:} {\rm Reviews in
Modern Astronomy {\bf 14},
15--51 (2001)}}

\vspace*{1.2cm}

\begin{center}

{\sl Ludwig Biermann Award Lecture}\\[0.7cm]

{\Large\bf
The Silent Majority\\[0.2cm]
Jets and Radio Cores\\[0.1cm]
from Low-Luminosity Black Holes}\\[0.7cm]

Heino Falcke\\[0.17cm]
Max-Planck-Institut f\"ur Radioastronomie\\
Auf dem H\"ugel 69, 53121 Bonn, Germany\\
{\tt hfalcke@mpifr-bonn.mpg.de,
http://www.mpifr-bonn.mpg.de/staff/hfalcke}
\end{center}

\vspace{0.5cm}

\begin{abstract}
\noindent{\it
They are weak, they are small, and they are often overlooked, but they
are numerous and an ubiquitous sign of accreting black holes: compact
radio cores and jets in low-power AGN.  Here I summarize our work
concerning these radio cores and jets in recent years, specifically
focusing on the large population of low-luminosity AGN. Special
attention is also given to Sgr A*, the supermassive black hole candidate at
the Galactic Center, whose radio properties are reviewed in more
detail. This source exhibits a submm-bump, possibly from an
ultra-compact region around the black hole which should allow imaging
of the event horizon of the black hole in the not too distant
future. A jet model is proposed which explains the basic feature of
Sgr A*: its slightly inverted radio spectrum, the submm-bump, the lack
of extended emission, and the X-ray emission.  This model also works
for famous sources like M81*, NGC4258, or GRS1915+105 based on the
argument that radio cores are jets whose emission can be scaled with
the accretion power over many orders of magnitude. This scaling is
corroborated by the detection of many Sgr A*-like radio cores in
nearby Low-Luminosity AGN (LLAGN), some of which show jet structures
on the VLBI (Very Long Baseline Interferometry) scale. These cores
confirm an AGN origin of about half of the known low-luminosity AGN
classified as LINERs and dwarf-Seyferts. It is argued that in fact
most of the compact radio emission at centimeter waves in LLAGN is
produced by a compact radio jet and not an Advection Dominated
Accretion Flow (ADAF). In general one can say that compact radio cores
are a genuine feature of AGN, allowing one to precisely pinpoint black
holes in many galaxies---not only in luminous quasars.}
\end{abstract}

\section{Introduction}\label{intro}
\subsection{The Vocal Minority}
\citeme{Falcke1998c}
One of the main subjects for radio astronomers has been the study of
extragalactic radio jets. When observed at a higher resolution, many
of the first radio sources discovered in the early years of radio
astronomy later turned out to be powerful, collimated flows of
relativistic plasma (called jets) which were ejected from the nucleus
of giant elliptical galaxies. These structures can reach sizes of
several million light years and hence extend well beyond their host
galaxies into the vastness of intergalactic space. This relative
isolation is ideal for studying the physics of astrophysical plasma flows
in great detail (see e.g.,
\citeNP{BridlePerley1984,BridleHoughLonsdale1994,KleinMackStrom1994,MartiMuellerFont1997})
and allows us to make some estimates of the properties of the IGM
(inter-galactic medium, e.g., \citeNP{SubrahmanyanSaripalli1993}).

An even more important aspect of radio jets, however, is that they are
the largest and most visible sign -- literally the "smoking gun" -- of
Active Galactic Nuclei (AGN). The standard model of an AGN consists of
a supermassive black hole at the center of a galaxy \cite{Rees1984}
that accretes matter via an accretion disk
\cite{vonWeizsaecker1948,Luest1952,ShakuraSunyaev1973,Lynden-BellPringle1974}.
The inflow of matter in the potential well leads to an enormous energy
production that is released through infrared (IR), optical,
ultra-violet (UV), and X-ray emission. In a fraction of sources one
also sees very strong $\gamma$-ray and TeV emission. Some of the
energy is then funneled into the relativistic radio jet along the
rotation axis of the disk. It was, in fact, the strong radio emission
from these jets which first led to the discovery of quasars (3C273,
\citeNP{HazardMackeyShimmins1963,Schmidt1963}).

Jets have therefore been studied with great interest over many years
and in this time a huge zoo of different jet species has emerged,
e.g. FR I\,\&\,II radio galaxies, compact steep spectrum (CSS) and
Gigahertz peaked spectrum (GPS) sources, blazars and BL Lacs (e.g.,
\citeNP{UrryPadovani1995}).  The main reason why all these radio
galaxies and radio-loud quasars have been studied in such detail so
far is their large radio flux densities of 100 mJy up to several tens
of Jy\footnote{1 Jy = 1 Jansky = 1 Watt m$^{-2}$ Hz$^{-1}$}, which
makes them easily accessible with current technology. On the other
hand, it was noted early on that the majority of quasars and AGN in
the universe are {\it not} radio-loud
\cite{StrittmatterHillPauliny-Toth1980,KellermannSramekSchmidt1989}
and have in fact a rather low radio flux, despite having very similar
optical properties compared to radio-loud quasars\footnote{This does,
however, not exclude that radio-quiet quasars have relativistic jets
as well (see
\citeNP{MillerRawlingsSaunders1993,FalckeSherwoodPatnaik1996,BrunthalerFalckeBower2000}).}. Moreover, classical radio galaxies and radio-loud
 quasars are among
the most luminous AGN we know, corresponding to black holes with the
largest masses and the largest accretion rates. Hence, when we discuss
the properties of relativistic jets in AGN, we usually tend to think
exclusively about a relatively small but vocal group of sources.  Is
this the whole universe, or just the tip of the iceberg?  Most likely
there are many more, weaker black holes and jets in the universe.

\subsection{The Silent Majority}
\citeme{FalckeNagarWilson2000}

The evidence for supermassive black holes in the nuclei of most
galaxies has become much stronger recently. Some of the best cases are
the Milky Way \cite{EckartGenzel1997}, NGC~4258
\cite{MiyoshiMoranHerrnstein1995}, and a number of other nearby
galaxies
\cite{FaberTremaineAjhar1997,RichstoneBenderBower1998,MagorrianTremaineRichstone1998}
where convincing dynamical evidence for black holes exists. Hence, the
basic powerhouse for an AGN -- the black hole -- is built into almost
every galaxy, but compared to quasars and radio galaxies there is a
huge range in power output between the most luminous quasars and barely
active galaxies like the Milky Way.

\begin{figure}
\centerline{\psfig{figure=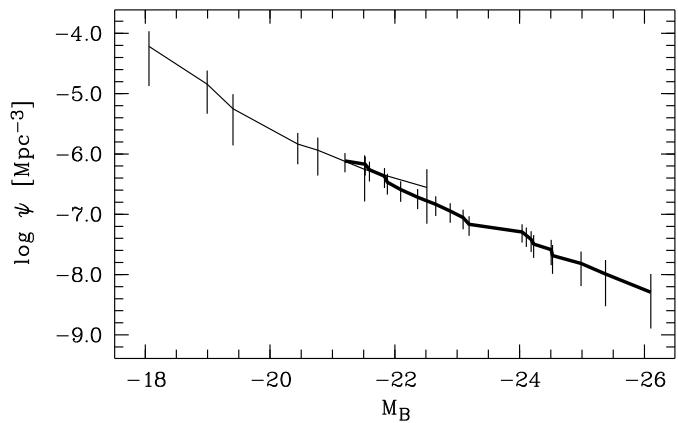,width=0.75\textwidth,bbllx=2.5cm,bblly=3.0cm,bburx=12cm,bbury=9.1cm,clip=}}
\caption{\label{llf}Local luminosity function (number density as a
function of absolute magnitude) of $z<0.3$ quasars and Seyferts. The
number of AGN decreases with a power-law index of $\alpha\simeq-2.2$ as
the luminosity increases (from K\"ohler et al.~1997).}
\end{figure}

For example in a spectroscopic survey of 486 nearby bright galaxies,
\citeN{HoFilippenkoSargent1997a}
found that a large fraction of these galaxies have optical
emission-line spectra. Roughly one third of the galaxies surveyed
showed spectra usually attributed to active galactic nuclei. The
energy output of these systems, is $10^{-6}-10^{-3}$ times lower than
in typical quasars \cite{Ho1999}. Consequently, these galaxies are
called low-luminosity AGN (LLAGN).  The large fraction of LLAGN
already indicates that the number of AGN increases with decreasing
luminosity. This is just the continuation of a trend that has been
found already in studies of the luminosity function of quasars and
Seyfert galaxies, namely a power-law distribution of AGN as a function
of luminosity with an index $\alpha\simeq-2.2$
(e.g.,~\citeNP{KoehlerGrooteReimers1997}, see Fig.~\ref{llf}). As in
real life, the majority of the entire population is rather quiet. To
get a complete view of the astrophysics of AGN and black holes one
therefore needs to look at this silent majority as well.

\nocite{KoehlerGrooteReimers1997}

The question of how the central engines in quasars and low-luminosity
AGN are related to each other and why they appear so different despite
being powered by the same type of object is therefore of major
interest. For many nearby galaxies with low luminosity nuclear
emission-lines, it is not even clear whether they are powered by an
AGN or by star formation. This is especially true for Low Ionization
Nuclear Emission Region (LINER) galaxies
\cite{Heckman1980,HeckmanvanBreugelMiley1983}, some of which can be
explained in terms of aging starbursts
(e.g.,~\citeNP{FilippenkoTerlevich1992,Alonso-HerreroRiekeRieke2000}).

One of the best ways to probe the very inner parts of these engines is
to study the compact radio sources found in many AGN. Indeed, despite
their low optical luminosity, quite a few nearby galaxies have such
radio sources in their nuclei, prominent cases being the Milky Way
(Sgr A*), and the galaxies M~104 and M~81
\cite{BietenholzBartelRupen1996}. These radio sources resemble the
cores of radio-loud quasars, showing a very high brightness
temperature and a flat to inverted radio spectrum that extends up to
sub-millimeter (submm) wavelengths.

\subsection{Radio Cores, Jets, and Accretion}
\citeme{FalckeBiermann1999}
What are these compact radio cores and how are they related to the
AGN?  Early on in the discussion about the existence of black holes,
\citeN{Lynden-BellRees1971} suggested that they would be accompanied by
compact radio nuclei, detectable by Very Long Baseline Interferometry
(VLBI), and predicted such a source for the Galactic Center. Indeed,
this source (Sgr A*) was then discovered by
\citeN{BalickBrown1974} and it became clear in later years that
compact radio cores are indeed good evidence for the existence of an
AGN or a black holes in a galaxy. For luminous radio galaxies and
radio-loud quasars the basic nature of these compact radio nuclei has
been clarified in the meantime through extensive and detailed VLBI
observations (see Zensus 1997 for a review) as being the inner regions
of relativistic jets emanating from the nucleus.

Despite this progress, a number of important questions remain when
looking back at the initial discussion. First of all, it is unclear
whether there indeed is a direct link between compact radio cores and
AGN, i.e.~whether compact radio cores and jets are just an accidental
by-product of black hole activity or a necessary
consequence. Secondly, for the lesser studied, low-luminosity AGN
the jet nature of compact radio nuclei has not yet been
established beyond any doubt, leaving the question open whether in
fact a compact radio core in a low-luminosity AGN is the same as in a
high-luminosity AGN, i.e.~a quasar.

Such a similarity was exactly the claim made by
\citeN{FalckeBiermann1995}, stimulated by the seminal paper by
\citeN{RawlingsSaunders1991}, where it was proposed that accretion
disks and jets form symbiotic systems. A scaling law was proposed
which connects high-power and lower-power accretion disks and their
associated radio jets (cores). Alternatively it was argued that radio
cores in LLAGN could be due to the accretion flow itself if it becomes
advection dominated
\cite{NarayanYi1994,NarayanYi1995a,NarayanYi1995b}. This idea was
applied to a range of objects including the Galactic Center
(\citeNP{NarayanYiMahadevan1996}; see also \citeNP{Rees1982} and
\citeNP{Melia1994}).

The purpose of this work is therefore to describe a comprehensive
study of the radio emission from AGN operating at powers less than
those of typical quasars and to clarify their nature. By going to
lower powers we want to understand how universal the central engine
really is. Are radio cores in LLAGN different from those in quasars?
How important is the formation of jets at lower power?  Are there
classes of sources where no jet is found? What happens with the jets
in quasars if the power of the engine becomes less and less: Will the
jets die completely, implying that accretion near the Eddington limit
is required for the jet formation, or will the jet just become
proportionally weaker, implying that jet formation is an integral part
of accretion physics and independent of the exact nature of the
accretion disk itself?

\section{The Jet-Disk Symbiosis Model}\label{symbiosis}
\citeme{Falcke1996f}

\subsection{Jet-Disk Coupling}
\citeme{Falcke1996f} In order to quantify the radio emission we expect
from a radio jet close to the nucleus we will make a few simple
assumptions and resort to the simple Ansatz that every jet is coupled
to an accretion disk. A coupled jet-disk system has to obey the same
conservation laws as all other physical systems, i.e.~at least energy
and mass conservation. We can express those constraints by specifying
that the total jet power $Q_{\rm jet}$ of the two oppositely directed
beams is a fraction $2q_{\rm j}<1$ of the accretion power $Q_{\rm
disk}=\dot M_{\rm disk}c^2$, the jet mass loss is a fraction $2q_{\rm
m}<1$ of the disk accretion rate $\dot M_{\rm disk}$, and the disk
luminosity is a fraction $q_{\rm l}<1$ of $Q_{\rm disk}$ for a standard optically thick disk ($q_{\rm
l}=0.05-0.3$ depending on the spin of the black hole).  The
dimensionless jet power $q_{\rm j}$ and mass loss rate $q_{\rm m}$ are
coupled by the relativistic Bernoulli equation
(\citeNP{FalckeBiermann1995}) for a jet/disk-system. For a large range
in parameter space the total jet energy is dominated by the kinetic
energy such that one has $\gamma_{\rm j}q_{\rm m}\simeq q_{\rm j}$, in
case the jet reaches its maximum sound speed, $c/\sqrt{3}$, the
internal energy becomes of equal importance and one has $2\gamma_{\rm
j}q_{\rm m}\simeq q_{\rm j}$ ('maximal jet'). The internal energy is
assumed to be dominated by the magnetic field, turbulence, and
relativistic particles. We will constrain the discussion here to the
most efficient type of jet where we have equipartition between the
relativistic particles and the magnetic field and also have
equipartition between the internal and kinetic energy (i.e.~bulk
motion) -- one can show (see
\citeNP{FalckeBiermann1995}) that other, less efficient models would
fail to explain the highly efficient radio-loud quasars.

Knowing the jet energetics, we can describe the longitudinal structure
of the jet at first by assuming a constant jet velocity (beyond a
certain point) and free expansion according to the maximal sound speed
($c_{\rm s}\la c/\sqrt{3}$). For such a jet, the equations become very
simple. The magnetic field is given by

\begin{equation}
B_{\rm j}=0.3\,G\;Z_{\rm pc}^{-1}\sqrt{q_{\rm
j/l}L_{46}}
\end{equation} 
and the particle number density is
\begin{equation}
n=11\,{\rm cm}^{-3} L_{46} q_{\rm j/l} Z_{\rm pc}^{-2}
\end{equation} 
(in the jet rest frame). Here $Z_{\rm pc}$ is the distance from the
origin in parsec (pc), $L_{\rm 46}$ is the disk luminosity in
$10^{46}$ erg/sec, $2q_{\rm j/l}=2q_{\rm j}/q_{\rm l}=Q_{\rm
jet}/L_{\rm disk}$ is the ratio between jet power (two cones) and disk
luminosity which is of the order 0.1--1
\cite{FalckeMalkanBiermann1995} and $\gamma_{\rm j,5}=\gamma_{\rm j}/5$
($\beta_{\rm j}\simeq1$). If one calculates the synchrotron spectrum
of such a jet, one obtains locally a self-absorbed spectrum that peaks
at

\begin{equation}
\nu_{\rm ssa}=20\,{\rm GHz}\;{\cal D}{\left(q_{\rm j/l}L_{46}\right)^{2/3}
\over Z_{\rm pc}}\,\left({\gamma_{\rm e,100} 
\over\gamma_{\rm j,5} \sin i}\right)^{1/3}.
\end{equation}
Integration over the whole jet yields a flat spectrum with a
monochromatic luminosity of

\begin{equation}\label{radioopt}
L_{\nu}={ 1.3\cdot 10^{33}}\,{{\rm erg}\over
{\rm s\, Hz}}\;\left({q_{\rm j/l} L_{46} }\right)^{17/12}
{\cal D}^{13/6}\sin i^{1/6} \gamma_{\rm
e,100}^{5/6} \gamma_{\rm j,5}^{11/6},
\end{equation}
where $\gamma_{\rm e,100}$ is the minimum {\it electron} Lorentz
factor divided by 100, and ${\cal D}$ is the {\it bulk} jet Doppler
factor. At a redshift of 0.5 this luminosity corresponds to an
un-boosted flux of $\sim100$ mJy. The brightness temperature of the jet
is

\begin{equation}
{T}_{\rm b}=1.2\cdot 10^{11}\, {\rm K}\; {\cal D}^{4/5}{\left({
{\gamma_{\rm e,100}}^2 q_{\rm j/l} L_{46} \over
\gamma_{\rm j,5}^2 \beta_{\rm j}}\right)^{1/12}\sin i^{5/6}}
\end{equation} 
which is almost independent of all parameters except the Doppler
factor. An important factor that governs the synchrotron emissivity
is, of course, the relativistic electron distribution, for which we
have assumed a power-law distribution with index $p=2$ and a ratio 100
between maximum and minimum electron Lorentz factor. As we are
discussing here the most efficient jet model we also assume that all
electrons are accelerated (i.e.~$x_{\rm e}=1$ in
\citeNP{FalckeBiermann1995}), hence the only remaining parameter is
the minimum Lorentz factor of the electron distribution $\gamma_{\rm
e,100}$ determining the total electron energy content. In order to
reach the magnetic field equipartition value, which is close to the
kinetic jet power governed by the protons, we have to require
$\gamma_{e,100}\sim1$.

Thus, using $L_{\rm disk}\sim10^{46}$ erg/sec (typical
optical/UV-luminosity) and $\gamma_{\rm j}\sim 5$ (a few times the
escape speed from a black hole), we can easily explain pc-scale radio
cores at cm-wavelengths, with brightness temperatures of $10^{11}$ K
and fluxes of 100 mJy and more without any need for a `cosmic
conspiracy' (e.g., \citeNP{CottonWittelsShapiro1980}).

\subsection{UV/Radio Correlation}
\citeme{Falcke1996f}
Now, we will have to validate some of our assumptions and test the
jet-disk coupling derived above. To stay on safe grounds we will in
this section concentrate on the well-studied quasars. For these
sources we have to estimate the disk luminosity as
precisely as possible and compare it to their radio cores. The best
studied quasar sample so far is the PG quasar sample
\cite{SchmidtGreen1983}. For most sources in this sample
\citeN{SunMalkan1989}, using optical and IUE data, fitted the UV bump
with accretion-disk models and a few more were available in the
archive \cite{FalckeMalkanBiermann1995}. There are also excellent
photometric \cite{NeugebauerGreenMatthews1987} and spectroscopic data
\cite{BorosonGreen1992} available. Unlike the broadband UV-bump
fits, which give $L_{\rm disk}$ directly, emission-lines and continuum
colors do not give a direct estimate for the bolometric UV luminosity
and $L_{\rm disk}$. We will need to calibrate those values to obtain
an equivalent UV bump luminosity using the sources which have a
complete set of data available. This yields

\begin{equation}
\lg (L_{\rm disk}/{\rm erg\;s}^{-1})=2.85+\lg (L_{\rm [OIII]}/{\rm erg\;s}^{-1}),\label{oiii2uv}
\end{equation}
\begin{equation}
\lg (L_{\rm disk}/{\rm erg\;s}^{-1})=2.1+\lg (L_{\rm H{\beta}}/{\rm erg\;s}^{-1}),
\end{equation}
\begin{equation}
\lg (L_{\rm disk}/{\rm erg\;s}^{-1})=-0.4 M_{\rm b}+35.90.
\end{equation}

Here $L_{\rm [OIII]}$ and $L_{\rm H{\beta}}$ are the luminosities in
the [OIII]$\lambda5007$ and H$\beta$ (broad) emission-lines, $M_{\rm
b}$ is the absolute blue continuum magnitude as used by
\citeN{BorosonGreen1992}. The scatter in this relation, $\sim0.5$ in
the log, is shown in \citeN{FalckeMalkanBiermann1995}.

\begin{figure}
\centerline{
\psfig{figure=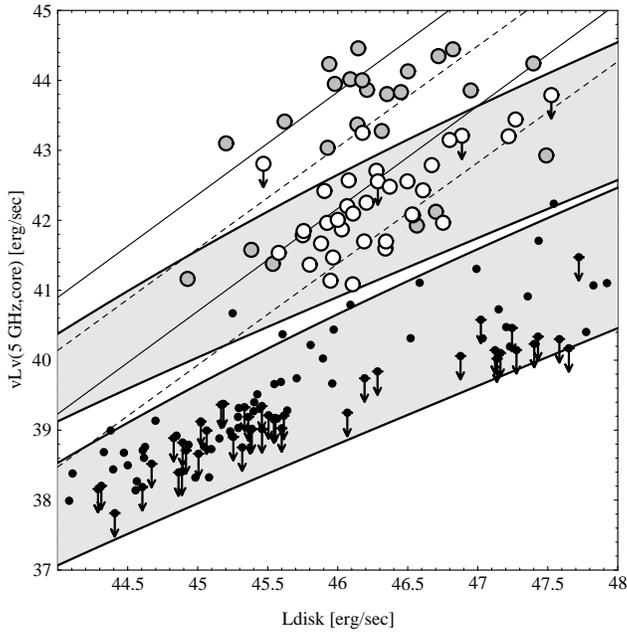,width=0.7\textwidth,bbllx=2.7cm,bblly=5.9cm,bburx=19.1cm,bbury=22.3cm}}
\caption[]{\label{uvradioplot-qso2}
Radio {\it core} luminosity vs.~disk (UV-bump) luminosity for quasars
(including a complete optical and a radio-selected sample). The shaded
circles are core-dominated, flat-spectrum sources, open circles are
steep-spectrum (FR II type) sources, and filled points are radio-quiet
sources.  The shaded bands represent the radio-loud and radio-quiet
jet models where the width is determined by relativistic boosting. The
jet velocity evolves with luminosity as $\gamma_{\rm j}\beta_{\rm
j}\propto L^{0.1}$, as discussed in Falcke, Malkan, Biermann
(1995). The dashed lines represent the expected level of emission for
sources just within the boosting cone (i.e.~inclination $i=1/\gamma$)
and the isolated solid lines represent emission for $i=0^\circ$
inclination -- corresponding to the maximally boosted flux. The
position of flat-spectrum and steep-spectrum sources and the
ra\-dio-loud/radio-quiet separation can be naturally accounted for
with the coupled jet-disk model.}
\end{figure}

In Figure \ref{uvradioplot-qso2} we plot the radio core luminosity
against the accretion disk luminosity derived from the above relations
for a quasar sample. This demonstrates that the {\it cores} of these
quasars show the distribution expected within the jet-disk model,
assuming one has relativistic jets in radio-loud {\em and} radio-quiet
quasars operating at two basic levels of efficiency (see
\citeNP{FalckeBiermann1995} for a discussion of what these basic
levels could be). The spread in the radio luminosity distribution is
dominated by relativistic boosting and random inclination angles. As
expected, flat spectrum radio-loud quasars, with the exception of a
few radio-intermediate quasars (possibly boosted radio-quiet quasars,
\citeNP{MillerRawlingsSaunders1993,FalckeSherwoodPatnaik1996}), are found at the highest radio
luminosities expected for jets with small inclination angles. This
spread can be translated into characteristic bulk Lorentz factors for
the jets, yielding a range between $\gamma_{\rm j}=3-10$.

\subsection[Radio Cores in LLAGN: Free Jets]{Radio Cores in LLAGN: Free Jet with Pressure Gradient}
\citeme{Falcke1996a}
Radio cores in quasars are quite well-studied and therefore one can
tolerate a number of free parameters in the jet model which are
constrained by additional observations. One of these parameters is the
jet speed which is generally believed to be in the range $\gamma_{\rm
jet}\sim5-10$ based on the observation of superluminal motion. In
LLAGN the situation is different and we have no direct evidence
whether they are highly-, moderately-, or sub-relativistic. It is
therefore useful to find a self-consistent description of the velocity
field in these radio cores mainly based on first principles. One
possibly important inconsistency of the basic jet model outlined
above, which is based on
\citeN{BlandfordKonigl1979}, is that it ignores the effects of any
pressure gradients: one can show that a self-consistent treatment
implies a weak acceleration of the bulk jet flow due to its
longitudinal pressure gradient. This cannot only naturally explain why
the radio spectra of some LLAGN radio cores are slightly inverted
(e.g., Sgr A* \& M81*; see
\citeNP{DuschlLesch1994,ReuterLesch1996,FalckeGossMatsuo1998}) rather
than just flat, it also provides one with a natural velocity field for
the jet, taking away $\gamma_{\rm jet}$ as a free parameter.

All these simplifications lead to the following expressions for the
observed flux density and angular size of a radio core observed at a
frequency $\nu$ as a function of jet power. For a source at a distance
$D$, with black hole mass $M_\bullet$, size of nozzle region $Z_{\rm
nozz}$ (in $R_{\rm g}=G M_\bullet/c^2$), jet power $Q_{\rm jet}$,
inclination angle $i$, and characteristic electron Lorentz factor
$\gamma_{\rm e}$\footnote{To simplify the equations we have defined
$\gamma_{\rm e}=\gamma_{\rm e,0}\left({Q_{\rm jet}\over
10^{39}\mbox{erg/sec}}\right)^{0.09} $} the observed flux density
spectrum is given as \cite{FalckeBiermann1999}

\begin{eqnarray}\label{simpleflux}
S_{\nu}&&=10^{2.06\cdot\xi_0}\;{\mbox mJy}\;\left({Q_{\rm jet}\over10^{39} \mbox{erg/sec}}\right)^{1.27\cdot\xi_1}
\nonumber\\&&\cdot
\left({D\over10{\rm kpc}}\right)^{-2}
\left({\nu\over8.5 {\rm GHz}}\right)^{0.20\cdot\xi_2}
\left({M_\bullet\over33 M_\odot}{Z_{\rm nozz}\over10 R_{\rm g}}\right)^{0.20\cdot\xi_2}
\nonumber\\&&\cdot
\left(3.9\cdot\xi_3\left(\gamma_{\rm e,0}\over200\right)^{-1.4\cdot\xi_4}
-2.9\cdot\xi_5\left(\gamma_{\rm e,0}\over200\right)^{-1.89\cdot\xi_6}\right),
\end{eqnarray}
with the correction factors $\xi_{0-6}$ depending on the
inclination angle $i$ (in radians):

\begin{eqnarray}
\xi_0&=&2.38 - 1.90\,i + 0.520\,{i^2}\\
\xi_1&=&1.12 - 0.19\,i + 0.067\,{i^2} \\
\xi_2&=&-0.155 + 1.79\,i - 0.634\,{i^2} \\
\xi_3&=&0.33 + 0.60\,i + 0.045\,{i^2} \\
\xi_4&=&0.68 + 0.50\,i - 0.177\,{i^2} \\
\xi_5&=&0.09 + 0.80\,i + 0.103\,{i^2}\\
\xi_6&=&1.19 - 0.29\,i + 0.101\,{i^2}.
\end{eqnarray}
Likewise, the characteristic angular size scale of the emission region
is given by

\begin{eqnarray}\label{simplesize}
\Phi_{\rm jet}&=&1.36\cdot\chi_0\,\mbox{mas}\,\sin{i}
\nonumber\\&\cdot&
\left(\gamma_{\rm e,0}\over200\right)^{1.77\cdot\chi_1}
\left({D\over10{\rm kpc}}\right)^{-1}
\left({\nu\over8.5 {\rm GHz}}\right)^{-0.89\cdot\chi_1}
\nonumber\\&\cdot&
\left({Q_{\rm jet}\over10^{39} \mbox{erg/sec}}\right)^{0.60\cdot\chi_1}
\left({M_\bullet\over33 M_\odot}{Z_{\rm nozz}\over10 R_{\rm g}}\right)^{0.11\cdot\chi_2},
\end{eqnarray}
with the correction factors

\begin{eqnarray}
\chi_0&=& 4.01 - 5.65\,i + 3.40\,{i^2} - 0.76\,{i^3}\\
\chi_1&=& 1.16 - 0.34\,i + 0.24\,{i^2} - 0.059\,{i^3}\\
\chi_2&=& -0.238 + 2.63\,i - 1.85\,{i^2} + 0.459\,{i^3},
\end{eqnarray}
where again the inclination angle $i$ is in radians. We point out that
in this model the characteristic size scale of the core region is
actually equivalent to the offset of the radio core from the
black hole. This does not exclude the existence of emission in
components further down the jet, which might be caused by shocks or
other processes.

\begin{figure}[t]
\centerline{\mbox{\psfig{figure=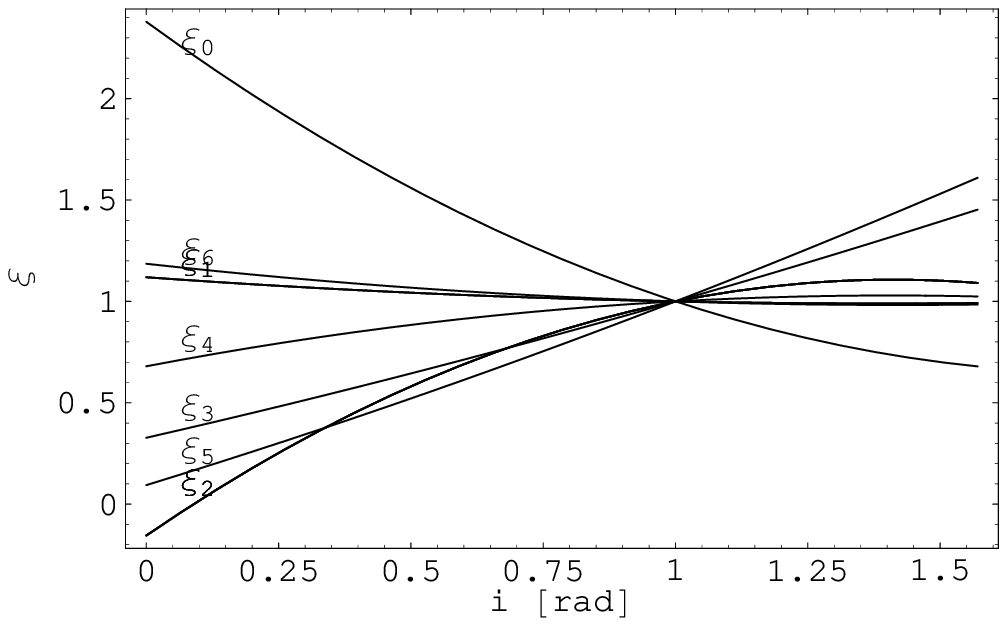,width=0.5\textwidth,bbllx=3.25cm,bburx=13.5cm,bblly=20.2cm,bbury=27cm}\hfill\psfig{figure=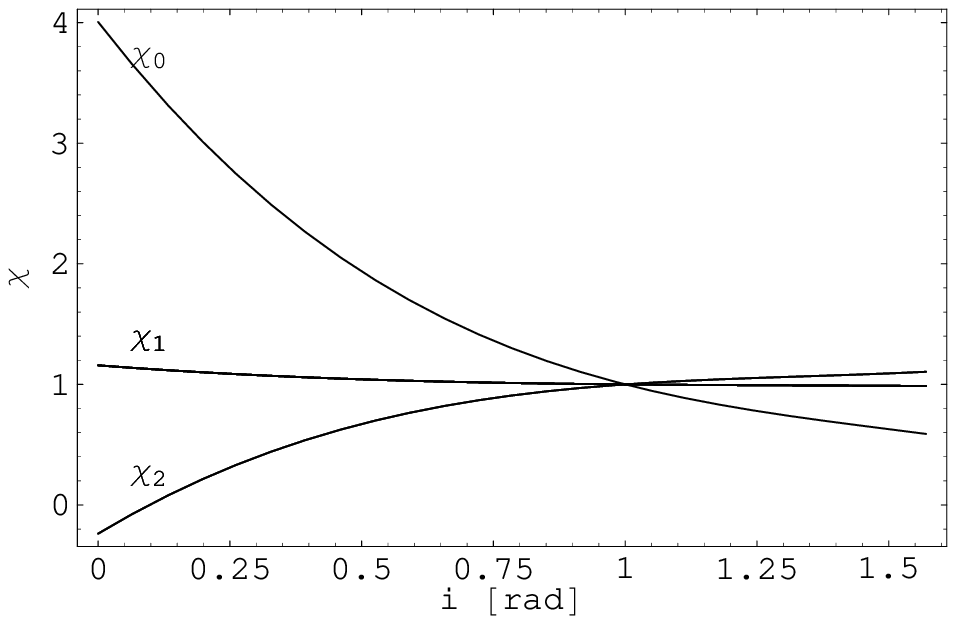,width=0.5\textwidth,bbllx=3.25cm,bburx=13.5cm,bblly=20.2cm,bbury=27cm}}}
\caption[]{Correction factors $\xi$ and $\chi$ for the exponents and
factors in Eqs.~\ref{simpleflux} \& \ref{simplesize} as a function of
inclination angle $i$ in radians, where $i=0$ corresponds to face-on
orientation. Note, however, that for $i\la10^\circ$ most
approximations fail.}
\end{figure}

For determining $Q_{\rm jet}$ from the jet model, the flux and to some
degree the inclination angle (especially for small $i$) are most
important, while $\gamma_{\rm e}$ is mainly determined by the size of
the core. Consequently, the latter seems to be the most uncertain part
since it is often ambiguous how to define the core and its
characteristic size. For practical purposes we can therefore give a
yet more simplified formula where we fix $\gamma_{\rm e}$ at an
intermediate value of 300 and which can be used to very roughly
estimate the jet power of a LLAGN from its flux and presumed
inclination angle alone (all $\xi_n$ are unity at $i=1$ rad):

\begin{equation}\label{jetpower}
Q_{\rm jet}=10^{37.6}\,\mbox{erg/sec}\;
{\frac{{{0.024}^
       {{\frac{{{\xi }_0}}{{{\xi }_1}}}}}\,
     {{1.56}^
       {{\frac{{{\xi }_4}}{{{\xi }_1}}}}}}{0.024\cdot1.56}}
{\frac{\,
     {{{\left({D\over10{\rm kpc}}\right)}}^
       {{\frac{1.6}{{{\xi }_1}}}}}\,
     {{{\left({S_\nu\over{\rm mJy}}\right)}}^
       {{\frac{0.79}{{{\xi }_1}}}}}}{
     {{{\left({M_\bullet\over33 M_\odot}\right)}}^
       {{\frac{0.15\,{{\xi }_2}}
          {{{\xi }_1}}}}}\,
     {{{\left({\nu\over8.5 {\rm GHz}}\right)}}^
       {{\frac{0.15\,{{\xi }_2}}
          {{{\xi }_1}}}}}\,
     {{{{\xi }_3}}^
       {{\frac{0.79}{{{\xi }_1}}}}}}}
\end{equation}

\subsection{Application to Famous Sources}
\setlength{\tabcolsep}{1pt}
\begin{deluxetable}{lcccccccccc}
\scriptsize\tiny\tablecolumns{10}\tablewidth{\textwidth}
\tablecaption{PARAMETERS OF WELL-STUDIED JET-DISK SOURCES}
\tablehead{
\colhead{Source }&
\colhead{ $D$ }&
\colhead{ $i$ }&
\colhead{$\nu_{\rm obs}$ }&
\colhead{ $S_\nu$ }&
\colhead{ size }&
\colhead{ $M_\bullet$
}&
\colhead{$L_{\rm disk}$ }&
\colhead{ $Q_{\rm jet}$ }&
\colhead{ $\lg \gamma_{\rm e}$ }\\
\colhead{       }&
\colhead{     }&
\colhead{  }&
\colhead{[GHz]              }&
\colhead{ [mJy]   }&
\colhead{ [mas]}&
\colhead{ [$M_\odot$]}&
\colhead{ [erg/sec]      }&
\colhead{ [erg/sec]     }&
\colhead{}&
}
\startdata
GRS~1915+105&12 kpc& 70$^\circ$& 15 & 40& 2 & (33) & $10^{39}$ &
$10^{39.1}$ & 2.6 \\
NGC~4258*&7.3 Mpc& 82$^\circ$& 22 & 3 & 0.4& $4\cdot10^7$& $(10^{42})$ & $10^{41.7}$ & 2.8 \\
M81*     &3.3 Mpc& 35$^\circ$& 8.5& 100 & 0.5& $(10^6)$ & $10^{41.5}$
& $10^{41.8}$ &2.4\tablenotetext{}{\label{famoustab}Parameters for compact radio cores in three famous
sources. Columns 2-7 are observationally determined input parameters:
distance $D$, inclination angle $i$ of disk axis and jet to the line
of sight, observing frequency $\nu_{\rm obs}$, flux density of radio
core $S_\nu$, size of radio core, and black hole mass $M_\bullet$. The
inferred disk luminosity $L_{\rm disk}$ is not an input parameter here
and given in column 8 for comparison only. Uncertain values are given
in brackets, but since the black hole masses do not enter strongly the
uncertainties in the black hole mass for M81 and GRS~1915 are actually
irrelevant. Columns~9-10 are output parameters of the radio core
model: jet power $Q_{\rm jet}$ and characteristic electron Lorentz factor
$\gamma_{\rm e}$.}
\enddata
\end{deluxetable}

One can apply the previously derived relations to various sources and
try to explain their radio core properties and derive their
basic parameters such as the jet power. This was done for example in
\citeN{FalckeBiermann1999} as summarized here in Table
\ref{famoustab}. The main result is that in three very
different, well-observed sources, with very different sizes and
fluxes, we can explain the central core with a single model by just
scaling the jet power with the accretion rate.  The high minimum jet
power found for the radio cores justifies in hindsight the assumptions
made in earlier papers that jet and disk can be considered symbiotic
systems and that {}--{} at least in a few systems {}--{} the
assumption of $Q_{\rm jet}/L_{\rm disk}\sim1$ (or even larger) seems
appropriate. This also strengthens the picture that the jets are
produced in the inner region of an accretion disk, where a major
fraction of the dissipated energy is channeled into the jet
\cite{FalckeBiermann1995,DoneaBiermann1996}.

Another consequence from those radio cores is their amazing scale
invariance.  It seems that we can use the very same model for a
stellar mass black hole which is accreting near its Eddington limit
(GRS~1915+105) as well as for a supermassive black hole which is
presumably accreting at an extreme sub-Eddington rate (M81,
NGC~4258). Moreover, a very similar model works successfully for the
quasar sample discussed in the previous section, i.e.~supermassive
black holes near the Eddington limit. 
This would suggest that certain
properties of an accretion disk/flow, namely jet production, is very
insensitive towards changes in accretion rate or black hole mass, and
that the 'common engine' mechanism of black hole accretion and jet
formation, suggested by
\citeN{RawlingsSaunders1991}, may include a much larger range of AGN
than only quasars and radio galaxies. To check this model we will
concentrate in the next section on the central black hole of the Milky
Way.

\section{The Galactic Center -- Sgr A*}\label{SgrA*}
The least luminous AGN we can observe is the center of our own
galaxy. The ever increasing amount of detailed observations severely
constrain any accretion models and give crucial informations for the
nature of nuclear radio cores. We therefore should have an in-depth
look at Sgr A*, which is the unique 1 Jy flat spectrum radio point
source located at the center of the Galaxy (a more detailed review can
be found in \citeNP{MeliaFalcke2001}). Within the Sgr A complex (see
\citeNP{MezgerDuschlZylka1996} for a review) it is surrounded by a
thermal radio source Sgr A West and embedded in the non-thermal
``hypernova'' shell Sgr A East.  Due to its unusual appearance it has
long been speculated that this source is powered by a supermassive
black hole. Its mass was believed to be around
$M_\bullet\sim2\cdot10^6M_{\odot}$ (e.g., Genzel \& Townes 1987). In
recent years high-resolution imaging \cite{EckartGenzelHofmann1993}
and spectroscopy \cite{HallerRiekeRieke1996} in the near-infrared
(NIR) have made this case much stronger and the detection of proper
motions of stars around the Galactic Center has solidified this case
even further \cite{EckartGenzel1996,GhezKleinMorris1998} yielding an
enclosed mass of $3\cdot10^6 M_{\sun}$. The latest development in
this direction is that first signs of acceleration in the star's
motion have been found and that the determination of entire orbits
will be possible soon \cite{GhezMorrisBecklin2000b}.  This will make
the mass determination yet more precise.  Moreover, since the relative
position of the radio source Sgr~A* in the NIR frame has been well
determined \cite{MentenReidEckart1997} one can also now state that Sgr
A* is in the dynamical center of the central star cluster of the
Galaxy within $0\farcs1$ \cite{GhezKleinMorris1998}. The constraints
from the first accelerations found in the stellar proper motions
mentioned above seem to have pushed this limit even further,
indicating that the center of mass is within 50 milli-arcsecond of Sgr
A*. Corroborating evidence that this radio source is associated with
the large amount of dark matter comes from the fact that, in great
contrast to the surrounding stars, Sgr A* does not show any random
proper motion down to a limit of 20 km/sec
\cite{ReidReadheadVermeulen1999,BackerSramek1999a}, while it does show
the apparent linear motion on the sky expected from the sun's rotation
around the Galactic Center. The lower limit placed on the mass of Sgr
A* from these observations is around 1000 $M_{\sun}$ {}--{} much
larger than any stellar object or remnant we know of in the Galaxy.

The enormous increase in observational data obtained for Sgr A* in
recent years has enabled us to develop, compare, and constrain a
variety of models for the emission characteristics of this source.
Because of its relative proximity and further observational input to
come Sgr A* may therefore become one of the best laboratories for
studying supermassive black hole candidates and basic AGN physics at
the lowest powers. Hence it is worth to dedicate a separate section to
this radio source and briefly summarize what its radio properties are
and how we can model its emission.

\subsection{The Size of Sgr A*}
A problem in determining the size of Sgr A* is that its true structure
is washed out by scattering in the interstellar medium
\cite{DaviesWalshBooth1976,vanLangeveldeFrailCordes1992,Yusef-ZadehCottonWardle1994,LoShenZhao1998}
leading to a $\lambda^2$ dependence of the apparent size of Sgr A* as
a function of observed wavelength.  Nevertheless, the mm-to-submm size
of Sgr A* is constrained at least within one order of magnitude. From
the absence of refractive scintillation \citeN{GwinnDanenTran1991}
have argued that Sgr A* must be larger than $10^{12}$ cm at
$\lambda1.3$ and $\lambda0.8$ mm. \citeN{KrichbaumZensusWitzel1993}
obtained a source size for Sgr A* of $9.5\cdot10^{13}$cm at 43 GHz
with VLBI -- well above the expected scattering size as extrapolated
from lower frequencies. This claim was challenged by
\citeN{RogersDoelemanWright1994} who only got $2\cdot 10^{13}$ cm at
86 GHz consistent with the scattering size.
\citeN{KrichbaumZensusWitzel1993} also found an additional weak
component and a somewhat elongated source structure at 43 GHz not seen
by \citeN{BackerZensusKellermann1993}. The problems of interpreting
elongated source structures in Sgr A* with insufficient baseline
coverage was discussed recently by
\citeN{DoelemanRogersBacker1999}. The most recent results are by
\citeN{BowerBacker1998} who find a source size of $6\cdot10^{13}$ cm
at 43 GHz ($\lambda7$mm) -- a mere 2 $\sigma$ above the scattering
size, while \citeN{LoShenZhao1998} infer an elongated source in
North-South direction with a size of $5.6\cdot1.5\cdot 10^{13}$
cm. The latter result is a 4 $\sigma$ deviation from the scattering
size and would either confirm the basic results of
\citeN{KrichbaumZensusWitzel1993} of a jet-like structure or suffer
the same problems as the earlier observations. Finally, observations
at 86 ($\lambda3$mm) and 215 GHz ($\lambda1.4$mm) by
\citeN{KrichbaumGrahamWitzel1998} and \citeN{DoelemanShenRogers2001} demonstrate
that Sgr A* is compact at a scale around some $10^{13}$ cm at the
highest frequencies, i.e.~some 11 Schwarzschild radii for a
$3\cdot10^6M_\odot$ black hole. While the exact size of Sgr A* cannot
yet be stated with absolute certainty, the latest observations fuel
hopes that somewhere in the millimeter wave regime the intrinsic
source size will finally dominate over interstellar broadening.

\subsection{The Spectrum of Sgr~A*}\label{bump}
\citeme{Falcke1997b,FalckeGossMatsuo1998}
There is also some confusion in the current literature about the
actual spectrum of Sgr A*. \citeN{DuschlLesch1994} compiled an average
spectrum from the literature and claimed a $\nu^{1/3}$ spectrum
indicative of optically thin emission from mono-energetic electrons.
However, in early simultaneous multi-frequency VLA observations
\cite{WrightBacker1993} the actual spectrum was bumpy and the
spectral index varied between $\alpha=0.19-0.34$
($S_{\nu}\propto\nu^\alpha$). \citeN{MorrisSerabyn1996} published a
more recent spectrum that was a smooth power law with $\alpha=0.25$
below 100 GHz.  At low frequencies there is a possible turnover at or
below 1 GHz \cite{DaviesWalshBooth1976}.

\begin{figure}
\centerline{\psfig{figure=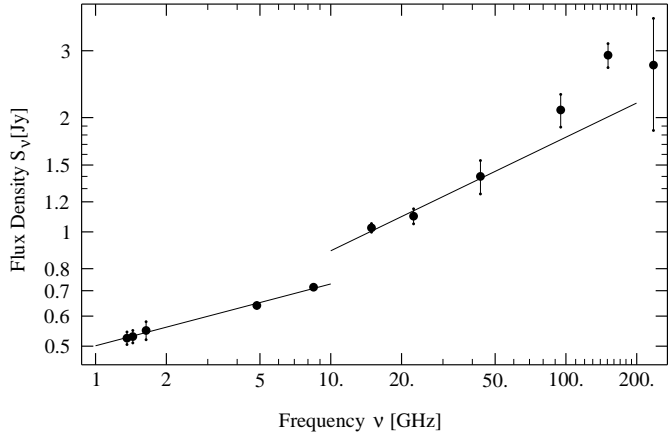,width=0.75\textwidth}}
\caption[]{\label{sgrspec}Quasi-simultaneous spectrum of Sgr~A* plotted as the
logarithm of the flux density vs.~the logarithm of the
frequency. Shown are the data averaged over the campaign period; flux
densities at neighboring frequencies were also combined from the
different mm-telescopes. Solid lines represent power law fits to the
low- and high-frequency VLA data.}
\end{figure}

Of greatest interest for the future debate is the
suggestion of a sub-millimeter (submm) bump in the spectrum
\cite{ZylkaMezgerLesch1992,ZylkaMezgerWard-Thompson1995}, since in all
models the highest frequencies correspond to the smallest spatial
scales. Simultaneous flux density measurement indeed seem to confirm
this excess (Fig.~\ref{sgrspec}; \citeNP{FalckeGossMatsuo1998}). In
the case of Sgr A* one expects the sub-millimeter emission to come
directly from the vicinity of the black hole \cite{Falcke1996b}.

The submm-bump causing this excess is, in fact, very well explained by
assuming the presence of a compact, self-absorbed synchrotron
component in Sgr~A*. As outlined in Falcke (1996b; see also
\citeNP{BeckertDuschl1997}) this submm component can be described in
its most simple minded form by four parameters: magnetic field $B$,
electron density $n_{\rm e}$, electron Lorentz factor $\gamma_{\rm
e}$, and volume $V=\frac{4}{3}\pi R^3$, using for simplicity a
one-temperature (i.e.~quasi mono-energetic) electron distribution and
the distance being set to 8.5 kpc.  On the other side we have three
measurable input parameters: the peak frequency $\nu_{\rm
max}\sim\nu_{\rm c}/3.5$ of a mono-energetic (or Maxwellian)
synchrotron spectrum (characteristic frequency $\nu_{\rm c}$), the
peak flux $S_{\nu_{\rm max}}$, and the synchrotron self-absorption
frequency $\nu_{\rm ssa}$. A fourth parameter can be gained if one
assumes that magnetic field and relativistic electrons are in
equipartition, i.e.~$B^2/8\pi=k^{-1} n_{\rm e} \gamma_{\rm e} m_{\rm
e} c^2$ with $k\sim1$. With this condition we obtain (averaged over
pitch angle) that

\begin{equation}\label{submmpars1}
\gamma_{\rm e}=118 \; k^{2/7}
\left({\nu_{\rm max}\over {\rm THz}}\right)^{5/17}
\left({\nu_{\rm ssa}\over 100 {\rm GHz}}\right)^{-5/17}
\left({F_{\nu_{\max}}\over 3.5 {\rm Jy}}\right)^{1/17},
\end{equation}

\begin{equation}
B=75\,{\rm G}\; k^{-4/17}
\left({\nu_{\rm max}\over {\rm THz}}\right)^{7/17}
\left({\nu_{\rm ssa}\over 100 {\rm GHz}}\right)^{10/17}
\left({F_{\nu_{\max}}\over 3.5 {\rm Jy}}\right)^{-2/17},
\end{equation}

\begin{equation}
n_{\rm e}=2\cdot10^6 {\rm cm^{3}}\;k^{7/17}
\left({\nu_{\rm max}\over {\rm THz}}\right)^{9/17}
\left({\nu_{\rm ssa}\over 100 {\rm GHz}}\right)^{25/17}
\left({F_{\nu_{\max}}\over 3.5 {\rm Jy}}\right)^{5/17},
\end{equation}

\begin{equation}\label{submmpars2}\label{sgrsizeeq}
R=1.5\cdot10^{12} {\rm cm}\;k^{-1/17}
\left({\nu_{\rm max}\over {\rm THz}}\right)^{-16/51}
\left({\nu_{\rm ssa}\over 100 {\rm GHz}}\right)^{-35/51}
\left({F_{\nu_{\max}}\over 3.5 {\rm Jy}}\right)^{8/17}
\end{equation}

Apparently the `non'-equipartition parameter $k$ enters only weakly
 and as long as one is not many orders of magnitude away from
 equipartition the parameters are basically fixed within a factor of a
 few and we can make  solid arguments about the possible source
 size of Sgr A* at submm wavelengths. As VLBI measurements are only
 available at longer wavelengths one could still postulate arbitrarily
 large submm source sizes.  However, if Sgr A*(submm) were optically
 thin and larger than $4\cdot10^{13}$ cm we should have seen the low
 frequency $\nu^{1/3}$ part of its spectrum with 3mm VLBI
 already. This could only be avoided if the submm component becomes
 optically thick below $\sim100$ GHz. As shown above this is possible
 only for a very compact source where the dimensions of Sgr A* at
 submm wavelengths are substantially {\it smaller} than at
 $\lambda$3mm. Consequently Sgr A* has to be of the same size or
 smaller at submm wavelengths than at $\lambda$3mm.
 
This size is consistent with the upper limits ($\sim$1~AU) from VLBI
and lower limits ($\sim10^{12}$cm) from scintillation experiments as
discussed above. In comparison we also note that the Schwarzschild
radius ($R_{\rm S}$) of the putative $3\cdot10^6 M_{\sun}$ black hole
in the Galactic Center is already $R_{\rm S}=0.9\cdot10^{12}$ cm and thus the
compact submm component should correspond to a region in the very
vicinity of the black hole. Most interesting is the possibility that
this region is directly affected by general relativistic effects, and
could for example be gravitationally amplified if the radiation is
intrinsically anisotropic (e.g., similar to \citeNP{Cunningham1975a}).
Finally, such a compact component with the parameters as in
Eq.~\ref{sgrsizeeq} would be very interesting for mm-VLBI, since the
black hole horizon could be imaged against the background of this
submm emission. This will be discussed in greater detail below
(Sec.~\ref{bhimage}).

Finally, as an important new window to Sgr A*, one needs to mention
that interesting new information on the polarization properties of Sgr
A* is available. In short one can say that Sgr A* stands out by having
relatively strong circular polarization but no detectable linear
polarization \cite{BowerBackerZhao1999,BowerFalckeBacker1999} below 43
GHz -- the situation at higher frequencies is still a bit unclear
\cite{BowerWrightBacker1999,AitkenGreavesChrysostomou2000}. Since
there is not enough space to discuss here the relevance of this
finding, one should just point out that this probably requires the
presence of rather cold ($T_{\rm e}\sim10^9$ K) electrons in addition
to the hot electrons required for producing the submm-bump ($T_{\rm
e}\sim10^{11}$K, Beckert et al., in prep.)

\subsection{Accretion Models -- An Overview}
\citeme{Falcke1996b} If we now want to go beyond a mere description of
Sgr A*, we have to ask how this source is powered and what the
underlying engine producing the radio and X-ray emission actually is?
One idea is that, if Sgr A* is a black hole, it should swallow some
fraction of the strong stellar winds seen in the Galactic Center through spherical
(Bondi-Hoyle) accretion.

The rate of infall depends only on the mass of the black hole and the
wind parameters.  The general validity of the
Bondi-Hoyle accretion (without angular momentum) under these
assumptions was demonstrated by 3D numerical calculations
\cite{RuffertMelia1994} and the main uncertainties are related to the
plasma physical effects associated with the infall. It is usually
assumed that the magnetic field in the accreted plasma is amplified by
compression up to a point where it reaches the equipartition value.
Beyond this point the excess magnetic field is assumed to be
dissipated and used to heat the plasma.  The electron temperature is
determined by the equilibrium between heating and cooling via
cyclo-synchrotron radiation where one has to consider two domains for
the solution of this problem: (1) hot electrons, where the typical
electron Lorentz factors are of the order 100-1000 and (2) warm
electrons, where the electron Lorentz factor is still close to unity.

The first domain is in a regime where synchrotron emission is
important and also very effective. This requires only low accretion
rates ($\dot M\sim10^{-10} M_{\odot}/{\rm yr}$) which seems to be in
contradiction with the naive expectations of the Bondi-Hoyle accretion
rate for a $3\cdot10^6 M_{\sun}$ black hole
\cite{Ozernoy1992,MastichiadisOzernoy1992}\footnote{These authors 
turned the argument around and used the reduced accretion rate to argue
for a lower mass black hole, something that seems to be excluded by
now}. The radio spectrum in such a configuration is mainly due to the
optically thin part of a quasi mono-energetic (or Maxwellian) electron
distribution and was calculated also by \citeN{DuschlLesch1994},
\citeN{BeckertDuschl1997}, and to some degree also 
used within the jet model (\citeNP{Falcke1996b}, but optically thick).

The second domain is in the transition regime between cyclotron and
synchrotron radiation, which is less effective than pure synchrotron
radiation and hence requires higher accretion rates ($\dot
M\sim10^{-4} M_{\odot}/{\rm yr}$;
\citeNP{MeliaJokipiiNarayanan1992,Melia1992a,Melia1994}) 
but also over-produces thermal X-rays (see below).

The big advantage of the wind-accretion approach would be that it,
firstly, appears unavoidable and, secondly, self-consistently ties
observable parameters and accretion rate to the mass of the central
object.

On the other hand there are several problems to be considered:
firstly, it is not at all clear that the wind-producing stars, as an
ensemble, have zero angular momentum, thus producing a non-zero
angular momentum of the winds to be accreted. This would diminish the
accretion rate. Hydrodynamical simulations show that the exact
distribution and velocity of stellar wind sources within the
observationally allowed range indeed can change the expected accretion
rate \cite{CokerMelia1997}. Another question is whether the stellar
winds are indeed isotropic. The only resolved stellar wind seen in the
Galactic Center so far, the wind of IRS 7
\cite{Yusef-ZadehMorris1991}, looks like a cometary tail pointing {\em 
away} from Sgr A*. This raises the question whether there is a wind
\cite{Chevalier1992} or expanding bubble pushing gas {\it out} of the
Galactic Center.

There also could be residual angular momentum in Sgr A* itself,
e.g.,~because of a fossil accretion disk which could catch the inflow
further out, filling a reservoir of rather dense matter instead of
directly feeding the black hole. The viscous time scales of such a
disk can be very long -- up to $10^7$ years \cite{FalckeHeinrich1994}.
However, detailed calculations of such a process
\cite{FalckeMelia1997,CokerMeliaFalcke1999} indicate that the impact
of the wind onto the fossil disk should produce a non-negligible IR
emission that is not seen.

An alternative to the models mentioned above was proposed by
\citeN{NarayanYiMahadevan1995} and
\citeN{NarayanMahadevanGrindlay1998}, who explain the discrepancy
between high accretion rate and low luminosity by the effects of an
advection dominated accretion flow (ADAF) or disk. In this model more
than $99.9\%$ of the energy is not radiated but transported through
the disk by advection and finally swallowed by the black hole. And, in
fact, it appears as if advection is non-negligible in many accretion
disks, but whether indeed such a high fraction of the energy is
transported by advection alone is not at all clear. One also has to
make sure that not a substantial fraction of the energy is released in
the inner parts of the disk and that the energy is swallowed quietly
by the black hole\footnote{But again, X-rays become more and more of a
problem.}. Even if an advection dominated accretion flow is not the
whole story, it may be an interesting part of it.

A major problem of the ADAF model is that it under-predicts the
cm-wave emission in the spectrum of Sgr A* by more than an order of
magnitude. In \citeN{NarayanMahadevanGrindlay1998} this is fixed by
artificially assuming a constant electron temperature in the accretion
flow over a particular range in radius $r$. The range was hand-picked
such that the cm-wave spectrum is fitted by a $\nu^{1/3}$ spectrum,
reminiscent of the spectra produced by \citeN{Ozernoy1992} and
\citeN{BeckertDuschl1997}. Another fix was proposed by
\citeN{Mahadevan1998} who added a power-law distribution of protons
producing high-energy pairs in hadronic interactions. The proton power
law would require shock acceleration in the accretion flow which
requires turbulent plasma waves
(e.g.,~\citeNP{SchlickeiserCampeanuLerche1993,Schneider1993}). It has
to be ensured that these waves do not couple to electrons as well,
otherwise this would lead to heating and shock acceleration of
electrons. In this case the basic assumption of a two-temperature
plasma, a fundamental assumption of ADAFs, would break down.

The latest information for Sgr A* comes from the possible detection of
Sgr A* at X-ray wavelengths with the satellite Chandra
\cite{BaganoffBautzBrandt2001}.  The first epoch data show a point
source at the location of Sgr A* with a rather low X-ray luminosity
around $0.5-1\cdot10^{34}$ erg s$^{-1}$ in the 0.5-10 keV band
(Baganoff et al., priv. comm.) {}--{} even lower than the earlier
ROSAT results \cite{PredehlTruemper1994}.  This new measurement provides a
crucial constraint for any model of radiative emission from Sgr A*
since the spectral index is apparently rather steep with a photon
index around 2.6 (Baganoff, priv. comm.). This seems to rule out
free-free emission from hot electrons as expected in Bondi-Hoyle or
ADAF models and requires those models to significantly lower their
accretion rate.

\subsection{The Jet Model for Sgr A*}\label{jetmodel}
\citeme{FalckeMarkoff2000}
In the following we will now concentrate on the jet model for Sgr A*
which allows one to self-consistently explain radio and X-ray emission
using the model outlined before. A more detailed description is given
in \citeN{FalckeMarkoff2000}.

Given an initial magnetic field $B_0$, a relativistic electron total
number density $n_0$ with a characteristic electron energy
$\gamma_{\rm e,0}m_{\rm e}c^2$, radius $r_0$ of the nozzle, and taking
only adiabatic cooling due to the longitudinal pressure gradient
(i.e.~$\propto {\cal M}^{\Gamma-1}$, where ${\cal M}$ is the Mach
number) and dilution by the lateral expansion into account, one can
determine the magnetic field $B(z)$, particle density $n(z)$, electron
Lorentz factor $\gamma_{\rm e,0}(z)$, and jet radius as a function of
the distance from the nozzle and then calculate the spectrum of the
jet in a straight forward way.

The basic parameters for the jet {}--{} $B_0$, $\gamma_{\rm e,0}$, and
$r_0$ {}--{} are fixed within a factor of a few by the location of the
submm-bump in the spectrum and which, in this model, is mainly
produced by emission from the nozzle \cite{Falcke1996b}. The steep
cut-off in the Sgr A* spectrum towards the IR further constrains the
electron distribution. It indicates the lack of a power-law tail of
high-energy electrons, which usually produces the optically thin
high-frequency emission seen, for example, in Blazars.

\begin{figure}
\centerline{\psfig{figure=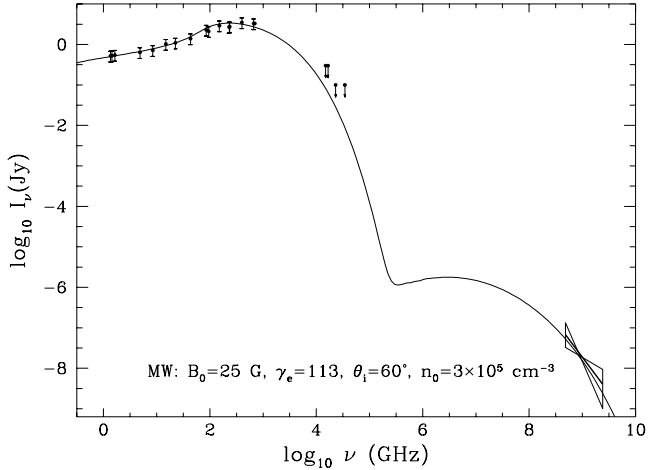,width=0.75\textwidth,angle=-90}}
\caption[]{\label{sgrx-pl}Broad-band spectrum of Sgr A* for a relativistic Maxwellian distribution of relativistic electrons in a jet. The
width of the nozzle is $r_0=4R_{\rm s}$ and $r_0=3 R_{\rm s}$
respectively, while its height is $z_0=3r_0$. The dots are the
simultaneous spectrum measured by Falcke et al.~(1998) with additional
high-frequency data from Serabyn et al.~(1997). In the hard X-rays we
show the possible detection of Sgr A* with Chandra (Baganoff et
al. 2001).}
\end{figure}

Figure \ref{sgrx-pl} shows a fit to the submm and cm radio spectrum
with the nozzle parameters given in the plot. We are plotting $F_\nu$
rather than $\nu F_\nu$ thus allowing one to better judge the quality
of the spectral fit at cm-waves. The nozzle component accounts for
most of the submm-bump, as well as the main Compton component
reproducing the X-ray emission, and the low-frequency radio spectrum
stems from the emission of the more distant parts along the
jet. Within the model, the slope of the cm-wave spectrum and the ratio
between cm and submm emission is mainly determined by the inclination
angle. The parameters we obtain for jet and nozzle are very close to
those used by \citeN{Falcke1996b}, \citeN{BeckertDuschl1997}, and
\citeN{FalckeBiermann1999}.  The jet-specific parameters appear
reasonable: the inclination angle is rather average, size of the
nozzle is a few times $R_s$, and the system is very close to
equipartition.

\nocite{FalckeGossMatsuo1998}\nocite{SerabynCarlstromLay1997}
\nocite{BaganoffBautzBrandt2001}

Since the X-ray emission is thought to be up-scattered submm-wave
emission, the predictions from this model are clear: we would expect
significant SSC emission in the softer X-ray band.  Given the
variability of the radio emission we also expect to see correlated
submm and X-ray variability.  Complete absence of X-ray variability
would argue against SSC emission giving a major contribution.

\subsubsection{VLBI Size and Extended Emission}
\begin{figure}
\centerline{\psfig{figure=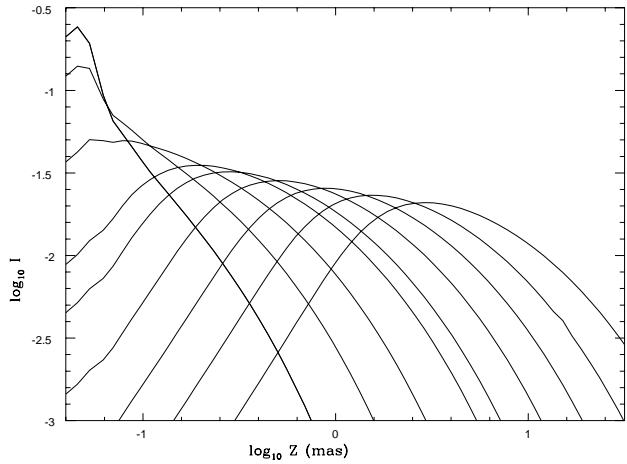,width=0.75\textwidth,angle=-90}}
\caption[]{\label{size.ps}Longitudinal cuts along the jet axis for
our model (Maxwellian electron distribution) at various frequencies
(220, 90, 43, 21, 15, 8.5, 4.8, 2.7, \& 1.4 GHz). Plotted are the
observed flux per segment in arbitrary units versus the observed
separation in milli-arcseconds (mas) from the black hole. Curves to
the left represent higher frequencies and the spikes in the left-most
flux distributions are due to the nozzle.}
\end{figure}

Possibly the most important constraints for any model are the VLBI
measurements of the size of Sgr A*. Since most models have a
stratified structure, the size of Sgr A* is expected to be a function
of frequency.  We note that for a given observing frequency the
emission in the jet model is highly concentrated to one spatial scale.
The emission from the jet at a particular frequency is self-absorbed
at small distances from the origin and cuts off at large distances
where the decreased magnetic field shifts the synchrotron cut-off
frequency below the observing frequency.  This is illustrated in
Fig.~\ref{size.ps}, where we show a longitudinal cut through the jet
based on our model. Each line corresponds to a different observing
frequency and one can see that the flux towards larger distances from
the core falls off rather steeply, i.e.~exponentially rather than with
a power law.

\nocite{LoShenZhao1998,KrichbaumGrahamWitzel1998}

\begin{figure}
\centerline{\psfig{figure=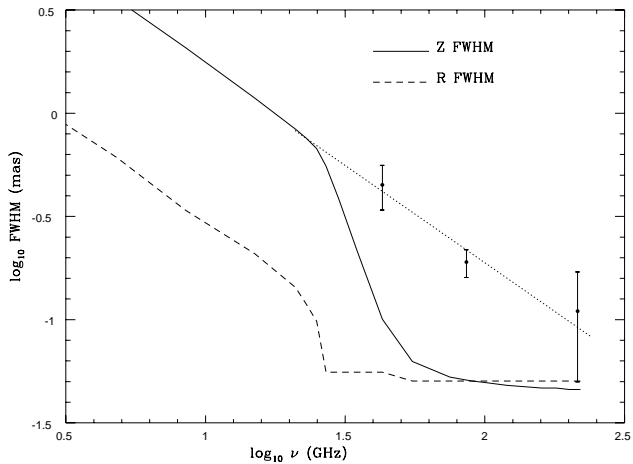,width=0.75\textwidth,angle=-90}}
\caption[]{\label{sgrx-size}FWHM of the major and minor axis of the
jet model as a function of frequency. The filled dots mark the FWHM as
measured by Lo et al.~(1998; 43 GHz) and Krichbaum et al.~(1998; 86
\& 215 GHz). At frequencies above 30 GHz one obtains a two component
structure with an increasingly stronger core (nozzle, solid dashed
line) and a fainter jet component (dotted line). This structure will
be largely washed out by interstellar scattering.}
\end{figure}

Thus, extended emission from the jet is highly (almost exponentially)
suppressed and the size of the detectable core will be a power law
$z\propto\nu^{-\aleph}$ with $\aleph$ in the range 0.9 to 1. This also
implies a shift of the location of the core with frequency. Figure
\ref{sgrx-size} compares the predicted full width at half maximum
(FWHM) of major and minor axis of the emission predicted by the jet
model with the constraints imposed by high-frequency VLBI
observations. Throughout the cm-wave range the emission basically
resembles one elliptical component decreasing in size with wavelength
and only at mm-waves (i.e.~above 30 GHz) an even more compact
core-component, the nozzle, appears. Hence, as long as the FWHM
predicted in the model is compatible with the observed values it will
be difficult to distinguish the Sgr A* source structure at one
frequency from a Gaussian VLBI component.

\nocite{LoShenZhao1998,KrichbaumGrahamWitzel1998}

\subsubsection{Conclusions from Applying the Jet Model}
The spectrum of Sgr A*, including the new X-ray observations from
Chandra, can be modeled entirely by emission from this jet alone. We
can also show that the radio emission satisfies all constraints
imposed by VLBI observations. This shows that the basic model
introduced by \citeN{FalckeMannheimBiermann1993} can provide a
detailed explanation of the Sgr A* radio and X-ray spectrum. It also
fits Sgr~A* within the frame work of compact radio cores discussed in
Sections~\ref{symbiosis} and \ref{llagn}.

One counter argument often heard in the context of modeling Sgr A* as
a jet is that one does not {\bf see} a jet.  However, in typical AGN
core-jet sources the extended jet structure is due to emission from an
optically thin power law. Here this extended emission is greatly
suppressed due to the steep cut-off in the electron spectrum, required
by the IR limits. This naturally can explain the compact (jet?)
structure as seen by \citeN{LoShenZhao1998} and
\citeN{DoelemanShenRogers2001}. Sgr A*, basically is a naked core
without much extended jet emission. The lack of optically thin
emission could also help to reduce the level of linear polarization in
Sgr A* compared to more powerful AGN
\cite{BowerBackerZhao1999,BowerFalckeBacker1999,BowerWrightBacker1999}
since the degree of polarization decreases to zero near the
self-absorption frequency due to radiation transfer effects.

The main assumption of the model is the presence of a nozzle close to
the central black hole collimating a relativistic plasma with
approximate equipartition between the magnetic field and relativistic
particles. The evolution of magnetic field and particle density is
calculated self-consistently and does not require additional
parameters beyond those fixed for the nozzle. The spectra we obtain
are therefore generic for collimated outflows from any accretion
flow {}--{} whether a magneto-hydrodynamical jet from a standard accretion
disk or an outflow from an ADAF {}--{} provided the accretion flow can
produce the required magnetic field, electron temperature, and density
near its inner edge. For an ADAF or Bondi-Hoyle type accretion the
presence of a jet near the black hole could thus aid those models in
accounting for the cm-wave radio emission, which is especially
difficult. The energy requirements to produce such a jet (see
\citeNP{FalckeBiermann1999}) are rather small compared to the power
available through accretion of nearby winds \cite{CokerMelia1997}.

It remains to be seen whether one can construct a self-consistent
model which couples an outflow as described here together with, for
example, an ADAF model (\citeNP{Yuan2000}). Particularly interesting
in this context is whether one could reproduce the unusual electron
distribution found in Sgr A*. As pointed out in \citeN{Falcke1996b}
the typical electron Lorentz factors required in the nozzle are close
to those expected from pair production in proton-proton (pp)
collisions. \citeN{Mahadevan1998} showed that an ADAF could in
principle provide the environment where pp-collisions could play an
important role even though in this paper a power-law distribution of
protons had to be artificially added. More detailed calculations in
this direction are being undertaken (Markoff et al., in prep.).

\subsection{Future Prospects - Imaging of the Event Horizon}\label{shadow}\label{bhimage}
The existence of a very compact nozzle radiating at mm- and
submm-wavelengths is intriguing since the size of Sgr~A* could be less
than 15 Schwarzschild radii (0.11 mas at 215 GHz) for a black hole
mass of $3\cdot10^6M_\odot$ at a distance of 8 kpc. Of all the known
black hole candidates, Sgr~A* is the source where the angular size on
the sky of its Schwarzschild radius is the largest. In fact, the
angular resolution of ground-based VLBI experiments now comes
interestingly close to the scale where significant general
relativistic effects are important.

We have considered an optically thin emission region with
frequency-independent emissivity around a black hole with arbitrary
spin \cite{FalckeMeliaAgol2000,FalckeMeliaAgol2000b}. The intensity
and structure of the emission region can be arbitrary as well, but we
choose a number of generic scenarios where we have either a spherical
distribution of the intensity scaling as a power-law with radius $r$
or a jet-like distribution (hollow cylinder). We also can allow for
various velocity fields, e.g.,~with rotation, inflow, or outflow.  The
appearance of the emission region for an observer at infinity, taking
all general relativistic effects into account, is then calculated
using a standard formalism
(e.g.,~\citeNP{Thorne1981,Viergutz1993a,JaroszynskiKurpiewski1997}).

\begin{figure}[b!] 
\centerline{\psfig{figure=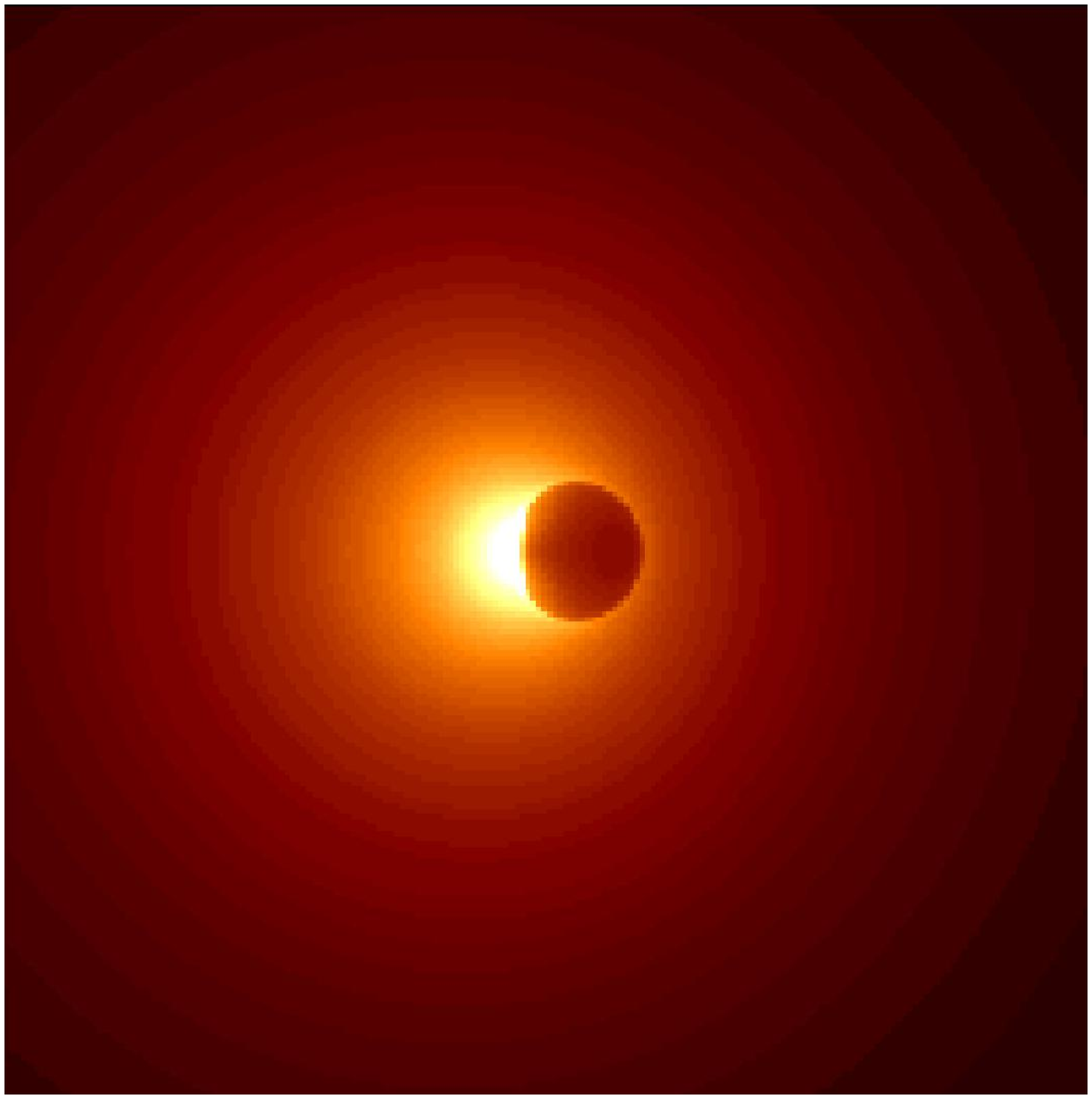,width=0.25\textwidth}\hskip-0.5mm\psfig{figure=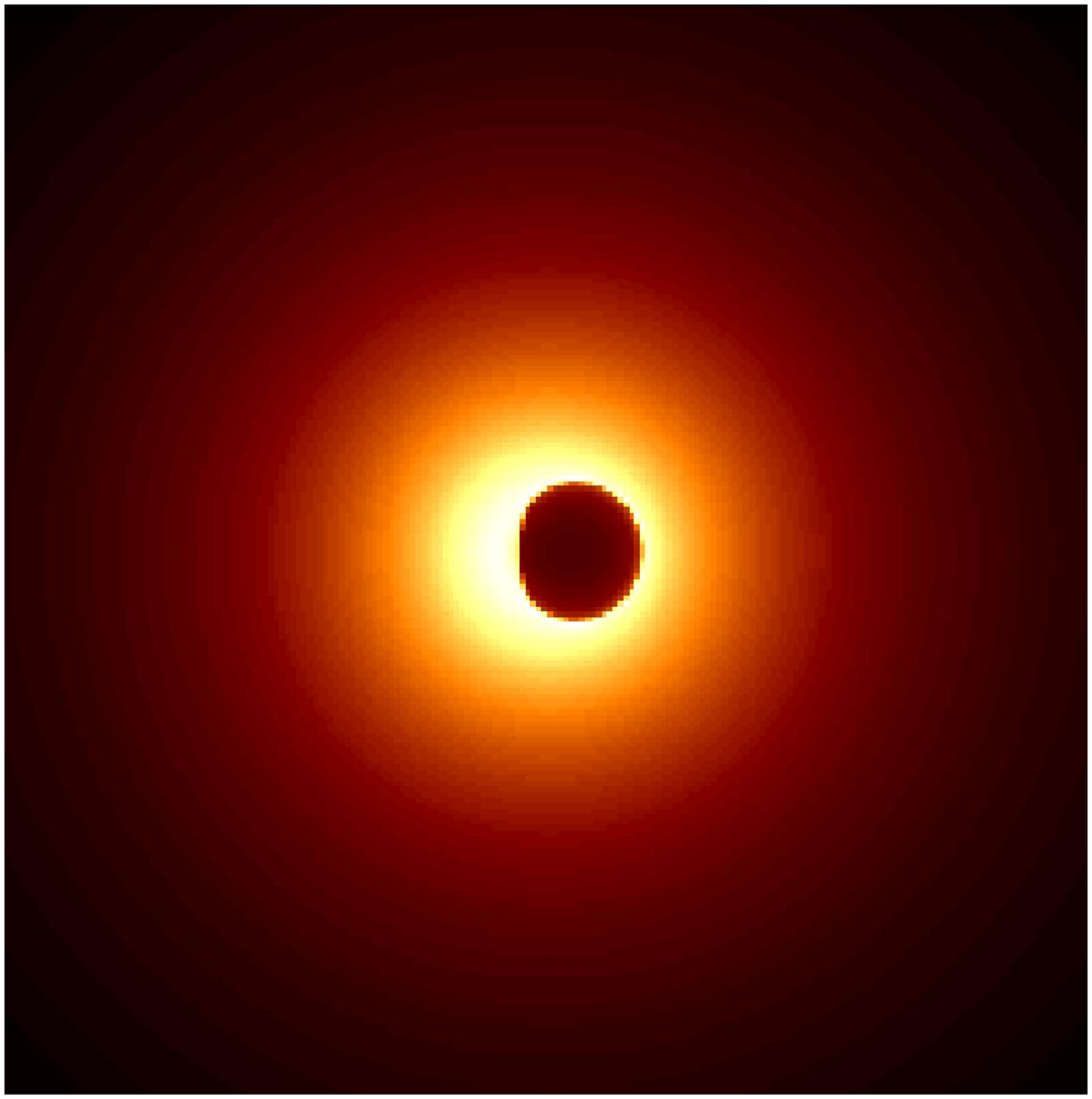,width=0.25\textwidth}\hskip-0.5mm\psfig{figure=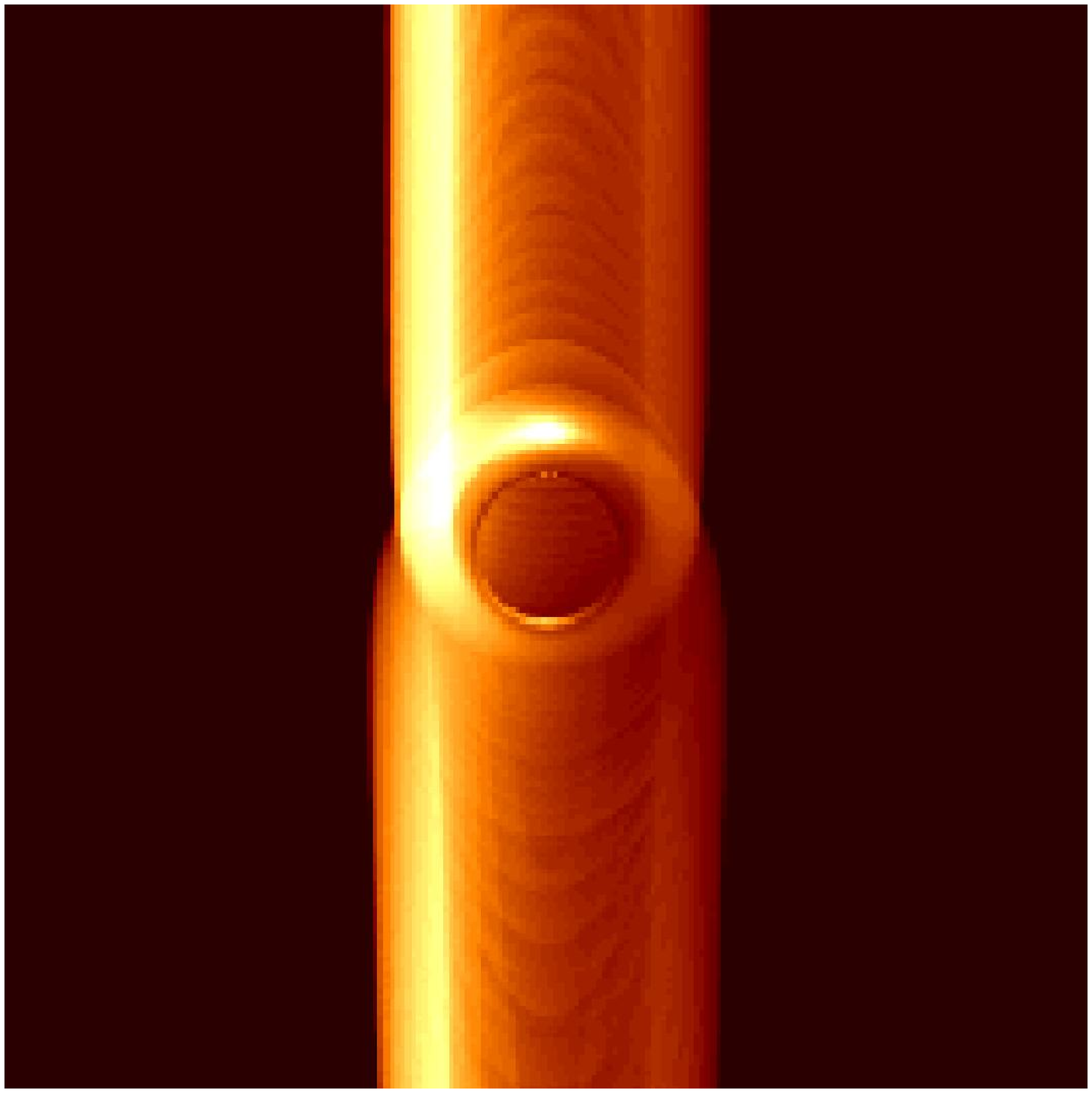,width=0.2512\textwidth}}
\caption{Images of the shadow of a black hole for rotating and
non-rotating black holes and for spherical and jet-like emission
models.}
\label{fig1}
\end{figure}

To test whether general relativistic effects would be visible we
convolved the resulting images from the ray-tracing calculations with
two Gaussian beams: one representing the scatter-broadening of the
image due to the interstellar material along our line-of-sight towards
the Galactic Center and one representing the finite resolution of VLBI
with 8000 kilometer baselines. The width of the former has a
$\nu^{-2}$ (e.g., \citeNP{LoShenZhao1998}) and the latter a $\nu^{-1}$
dependence.

Regardless of the exact emission model we use, we find a
characteristic structure in all models: a bright ring of emission with
a pronounced deficit of emission inside of that (Fig.~\ref{fig1}). We call the
deficit in the inner region the ``shadow'' of the black hole since it
is caused by the deficit of photons emitted near the black hole that
have disappeared into the event horizon or are bent away from our line
of sight. The circumference of the shadow is determined by the
`photon-orbit'---a theoretical orbit where photons can circle the
black hole an infinite number of times, but when perturbed may escape
to infinity \cite{Bardeen1973}. Interestingly, the size of this shadow is
much larger than the event horizon---due to gravitational
lensing---and is always of the order 10 $R_{\rm g}$ ($R_{\rm
g}=GM_\bullet/c^2$) for rotating and non-rotating black holes.

The exact intensity distribution of the bright ring depends
significantly on the nature of the emission region, however. A
rotating inflow would produce a slightly asymmetric ring due to
Doppler boosting of one side of the shells in Keplerian rotation. A
jet would look even more asymmetric since boosting due to rotation
plus fast outflow would enhance one quadrant of the ring
(Fig.~\ref{fig1}).

\begin{figure}
\centerline{\psfig{figure=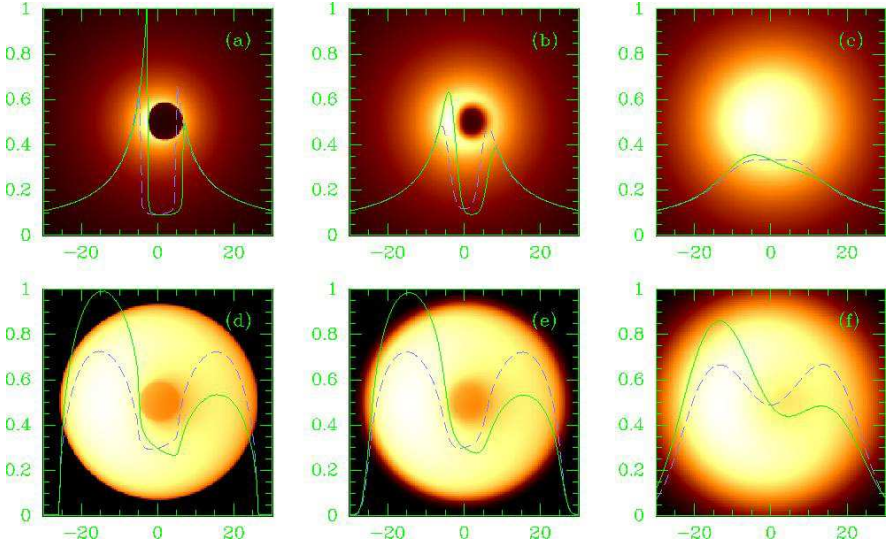,width=0.99\textwidth}}
\caption{The expected shadow of Sgr~A*. a) the ray-tracing
simulation; b) simulated VLBI image at $\lambda0.6$mm; c)
simulated VLBI image at $\lambda1.3$mm (see Falcke et al.~2000 for
details).}
\label{fig2}
\end{figure}

The relatively large size of the shadow is of particular interest for
Sgr~A*, since at a wavelength of around 1.3 mm the black hole shadow,
the scattering disk, and the possible resolution of mm-VLBI become
comparable. This is illustrated in Figure~\ref{fig2}. It is clear that
at wavelengths shortwards of $\lambda$1.3 mm the shadow could actually
be imaged with ground-based telescopes.

The possibility of seeing the effect of an event horizon is
tantalizing. The shadow of the black hole in the Galactic Center is
expected to have a diameter of $\sim30\,\mu$as. The highest resolution
so far achieved with VLBI is $\sim50\,\mu$as. To achieve the
additional improvement of a factor of two to three in resolution would
require extending mm-VLBI to submm wavelengths. While this is
difficult because of atmospheric effects, it is not technically
impossible.  Quite a number of submm-telescopes and arrays are
currently under construction or consideration that could be used for
such an experiment.

Another concern is whether the source itself could become an
obstacle. Clumpiness of the emission may not be a major source of
confusion because of the short rotation timescale of about 100
seconds. Optical depth effects could be a major problem. However,
currently available submm spectra of Sgr A* indicate a rather flat
spectrum with a turnover towards the infrared. Hence at some
wavelength between mm-radio and IR the source is bound to be optically
thin. A second pitfall is anisotropic beaming. In the jet model, for
example, one quadrant is amplified due to relativistic beaming, making
it more difficult to pick out the entire faint ring with low
resolution or low dynamic range observations.

Without a better understanding of spectrum, structure, and nature of
the emission it will be difficult to predict exactly at what
wavelength the shadow will be unambiguously detectable and how much
technical development still has to be done. In any case there is no
reason to think that imaging the shadow is in principal impossible and
any upcoming VLBI experiment at 1.3 mm and shorter wavelengths
involving Sgr~A* from now on could already show the first signs of the
event horizon.

\section{Low-Luminosity AGN}\label{llagn}
In Section \ref{symbiosis} we have discussed the general theory of
compact radio cores and the jet-disk symbiosis which was then applied
to Sgr A*. The conclusion was that compact radio cores are rather
scale invariant and only subject to changes in the accretion rate onto
the central black hole.  With the finding of black holes in many
nearby galaxies and the identification of a huge crowd of
low-luminosity AGN, many of them compiled in the spectroscopic atlas
by \citeN{HoFilippenkoSargent1995}, it was clear that from this theory
one would expect a large number of galaxies with compact radio cores
(see also
\citeNP{Perez-FournonBiermann1984}). These radio cores should bridge
the gap between Sgr A* and quasars.  The imperative then was to find
them in a significant number and compare their properties with more
luminous counter parts.

This comparison is also important since the question of how central
engines in high and low-luminosity AGN are related to each other and
why they appear so different despite being powered by the same type of
object is of major interest. For many nearby galaxies with
low-luminosity nuclear emission-lines, it is not even clear whether
they are indeed powered by an AGN or by star formation {}--{} despite
many of them being called LLAGN.

Earlier surveys have shown that E and S0 galaxies often have compact,
flat-spectrum radio sources in their nuclei
\cite{WrobelHeeschen1984,SleeSadlerReynolds1994}.  Some of the most
prominent flat-spectrum nuclear radio sources in nearby galaxies are
found in galaxies with LINER nuclear spectra
\cite{O'ConnellDressel1978}, but so far there has been no
comprehensive study of radio nuclei in a significant sample of LINER
galaxies, which make up the majority of galaxies with low-level
nuclear activity. We have, therefore, recently conducted a survey of
LINER galaxies with the VLA at 15 GHz (resolution $\sim$0\farcs15) and
the VLBA at 5 GHz (resolution $\sim$0\farcs002) to search for compact
radio emission (\citeNP{NagarFalckeWilson2000,FalckeNagarWilson2000}).
In the following sections the results of this search for ``siblings of
Sgr A*'' and their interpretation are presented.

\subsection[VLA Detection of Radio Cores]{Detection of Flat-Spectrum Radio Cores with the VLA}
\citeme{NagarFalckeWilson2000}
Observations were made of two samples
\cite{NagarFalckeWilson2000,FalckeNagarWilson2000}. All sources were
drawn from the extensive and sensitive spectroscopic study of the
complete, magnitude-limited sample of 486 nearby galaxies mentioned
above
\cite{HoFilippenkoSargent1995}, one third of which showed LINER-like
activity \cite{HoFilippenkoSargent1997a}. From these active galaxies
with a LINER spectrum a subsample (dubbed ``48 LINERs'' sample) of 48
bright sources was drawn with no well-defined selection criterion
other than that they had been observed with other telescopes as well,
e.g., ROSAT, the HST (UV imaging,
\citeNP{MaozFilippenkoHo1996,BarthHoFilippenko1998}), and the VLA at
1.4 and 8.4 GHz in A and B configuration
\cite{vanDykHo1998}.  The sample also included so-called transition
objects which have spectra intermediate between LINER and \ion{H}{2}
region galaxies. While the project was being conducted a few sources
in the original LINER sample were re-classified as low-luminosity
Seyfert galaxies. Transition objects were included because for these
objects it was not clear whether their emission-line spectrum is
produced by intense star formation or whether they are simply LINERs
with a very faint AGN.

In a second step we compiled a distance limited sample (dubbed ``96
LLAGN'' sample) from the \cite{HoFilippenkoSargent1995} atlas,
selecting all galaxies with LINER, Seyfert, and transition spectra
within 19 Mpc. This will reduce the effects of any bias that might
have come from the rather ill-defined selection criterion of the first
sample.  By including LINERs, Seyferts, and Transition objects we will
also be able to see possible differences in the detection rates
between the different types. By choosing a distance limited sample,
constrained to the very local universe, we also make sure to study the
faintest AGN we can find today.

Both samples were observed with the VLA at 15 GHz in its largest
configuration (A) providing maximal resolution, i.e.~up to 0\farcs15
corresponding to a linear scale of 14 pc at a distance of 19 Mpc. The
$5\sigma$ detection limit was around 1 mJy.

The rationale behind going to this setup is that the extended emission
from AGN is optically thin and steep-spectrum. Radio cores, on the
other hand -- if they are produced on scales very close to the AGN --
are optically thick, and compact at milli-arcsecond scales. This means
that by going to the configuration with the highest resolution one
will resolve out most of the extended emission and get a clearer view
on the compact structure. At higher frequencies this effect is even
stronger since the beam size is inversely proportional to the
frequency, reducing the extended flux per beam. The steep spectrum
will diminish the flux density of the extended emission even further.
The reason that one does not go to even higher frequencies is simply a
technical one, since the achievable sensitivity with the current
instrumentation of the VLA becomes increasingly worse at 22 and 43
GHz.

The results of the first survey for the 48 LINERs is that we detected
relatively many, namely 18 out of 48 galaxies (37\%). Split into the
different groups we detected only 1 out of 18 transition objects, but
6 out of 8 Seyferts and 11 out of 22 LINERs, yielding a combined
detection rate for Seyferts and LINERs of 57\% compared to only 6\%
for transition objects.

Using literature data (the VLA data at lower frequencies are still
being processed by another group) it was possible to determine whether
the detected cores should be considered as flat- or steep-spectrum
cores.  Out of the 18 detected cores only two apparently had a steep
radio spectrum. Hence, the 15 GHz VLA observations were indeed
properly designed to discriminate against steep-spectrum emission and
to pick out the flat-spectrum cores.


The high detection rate is confirmed by observations of the second
sample (Falcke et al., in prep.). Since this is a distance limited
sample selection effects should be minimized. The detection rate is 32
out of 95 galaxies (one galaxy was lacking data), i.e.~34\%. In the
sample we detected only 6 out of 38 transition galaxies (16\%), while
we detected 26 out of 57 Seyferts and LINERs (46\%). The large
majority of these detections are in fact confirmed as flat-spectrum
cores.

We can therefore conclude that galaxies with Seyfert and LINER spectra
are much more likely to contain flat-spectrum radio cores than
transition objects. The latter are therefore probably significantly
weaker AGN or are largely dominated by starbursts. On the other hand,
if one considers flat-spectrum cores as evidence for a black hole
powered AGN, we can conclude that at least about half of LINERs and
Seyferts are indeed genuine AGN.  If we can combine both samples we
find no significant difference in the detection rates between Seyferts
and LINERs.

However, one could still wonder whether the flat-spectrum cores are really
AGN related. In principle the flat spectrum could also be produced by
a giant free-free emission region in a star forming region. To exclude
this possibility one needs to go to even higher resolution and observe
the respective galaxies with VLBI. This will allow one to probe
brightness temperatures around $10^8$ K for the cores of interest here
and distinguish between the two cases.

\subsection[VLBA Detection of Radio Cores]{Detection of Flat-Spectrum Radio Cores with the VLBA}
\citeme{FalckeNagarWilson2000}

From our 48 LINERs sample (previous section), we selected all eleven
galaxies with both nuclear flux densities above 3.5 mJy at 15 GHz and
a flat spectrum ($\alpha>-0.5,\; S_\nu\propto\nu^\alpha$). The flux
density limit was chosen so that we could detect all sources with the
VLBA in snapshot mode in a single 12 hr observation if most of the 15
GHz emission were indeed compact on milli-arcsecond (mas) scales.  The
observations are more extensively described in
\citeN{FalckeNagarWilson2000} and 
indeed all sources were detected with the VLBA\footnote{The one source
for which the data reduction had initially failed was detected in a
subsequent observing run (Nagar et al., in prep.)}.

\begin{figure}
\centerline{\psfig{figure=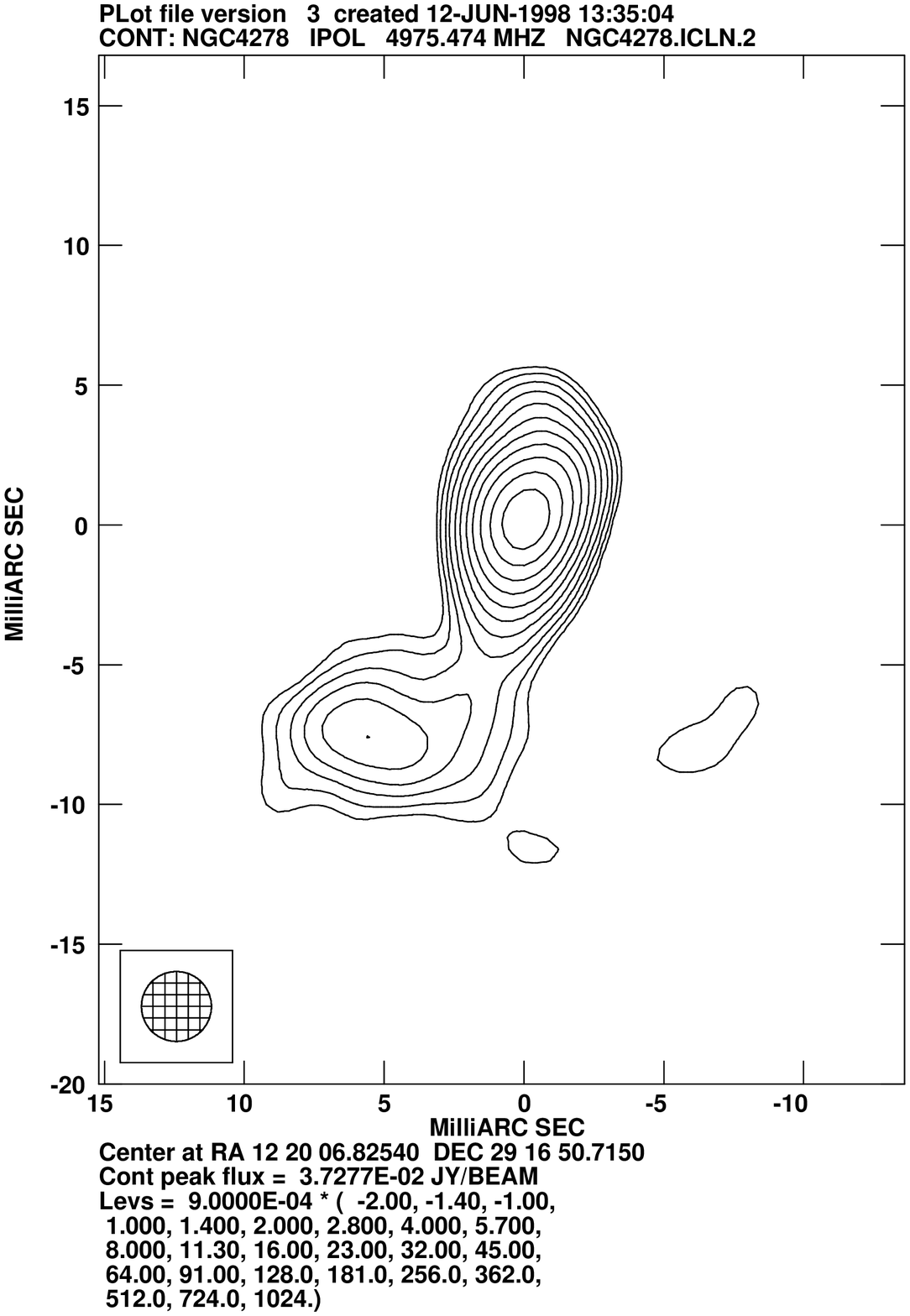,height=0.35\textwidth,bbllx=2.1cm,bburx=19.2cm,bblly=4.9cm,bbury=25.4cm,clip=}\quad\psfig{figure=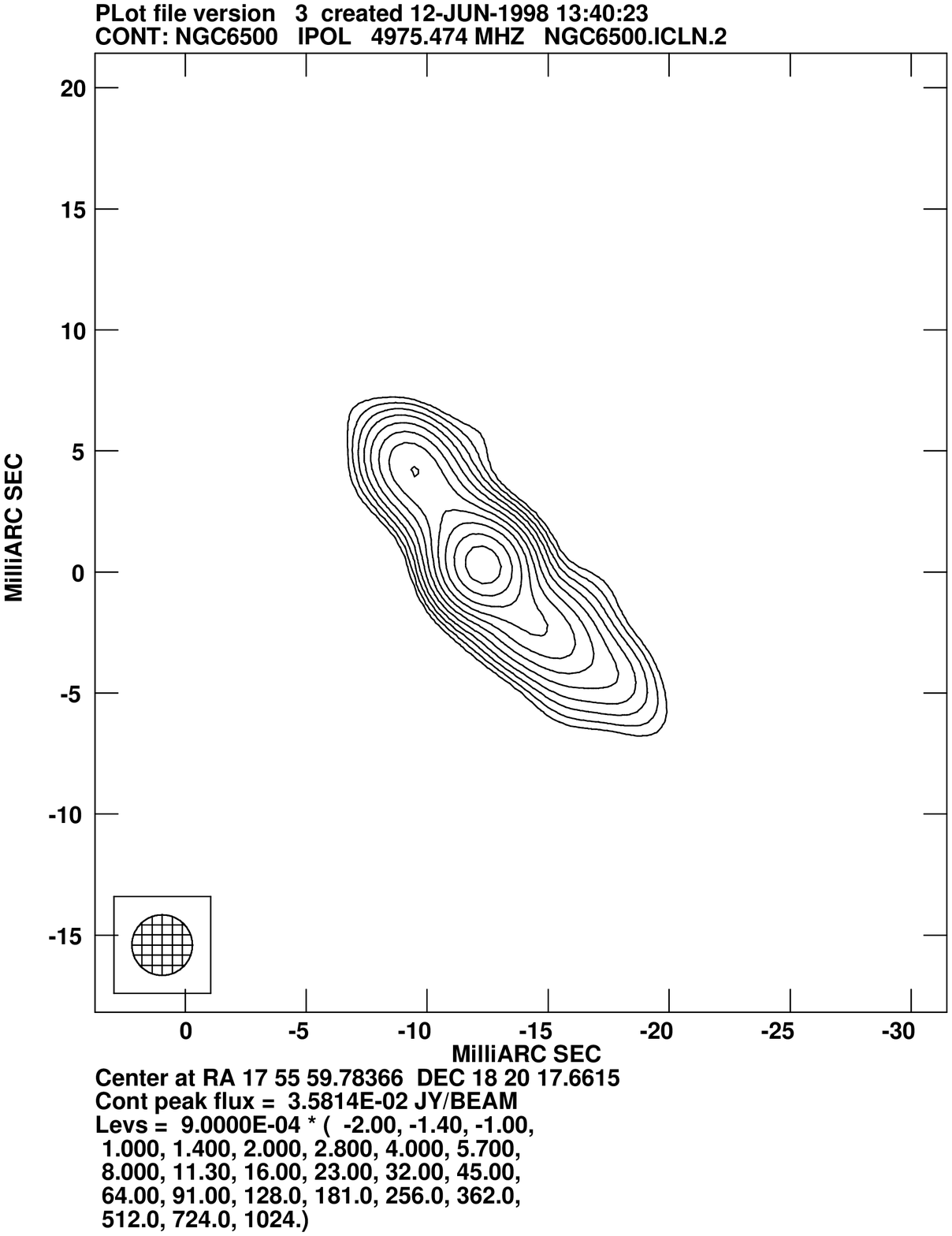,height=0.35\textwidth,bbllx=1.425cm,bburx=20.1cm,bblly=5.2cm,bbury=25.05cm,clip=}}
\caption[]{\label{LINVLBA}VLBA maps of NGC~4278 (left) and NGC~6500 (right). The beam is 2.5 milli-arcsecond and contours are integer powers
of $\sqrt{2}$, multiplied by the $\sim5\,\sigma$ noise level of 0.9
mJy.  The peak flux densities are 37.3 mJy and 35.8 mJy respectively.}
\end{figure}

The two brightest sources in our sample, NGC~4278 \& NGC~6500, for
which we have the largest dynamic range, show core plus jet structures
(Fig.~\ref{LINVLBA}).  The spectral indices of the whole sample range
from $\alpha=-0.5$ to $\alpha=0.2$, with an average
$\left<\alpha\right>=0.0\pm0.2$.  Using our beam size of 2.5
milli-arcsecond and the peak 5 GHz flux densities (Table 1 in
\citeNP{FalckeNagarWilson2000}), we find brightness temperatures in the
range $T=0.25-2.9\cdot10^8$ K for our sample, with an average
brightness temperature for all sources of $\left<T_{\rm
b}\right>=1.0\cdot10^8$ K. Since most of our sources are unresolved,
these values are usually lower limits.

Our result has a number of interesting implications. The presence of
high brightness temperature radio cores in our LINER sample confirms
the presence of AGN-like activity in these galaxies. It is unlikely
that the radio sources represent free-free emission, as has been
claimed for example in NGC~1068 \cite{GallimoreBaumO'Dea1997}, since a
much higher soft X-ray luminosity than is typically observed in
low-luminosity AGN would be expected.

On the other hand, the compact, flat-spectrum cores we have found are
similar to those typically produced in many AGN. Hence we can take the
presence of compact, non-thermal radio emission as good evidence for
the presence of an AGN in our galaxies.  The 100\% detection rate with
the VLBA, based on our selection of flat-spectrum cores found in a 15
GHz VLA survey, shows that for statistical purposes we could have
relied on the VLA alone for identification of these compact, high
brightness radio sources.  Hence, with 15 GHz VLA surveys of nearby
galaxies one has an efficient tool for identifying low-luminosity
AGN. This complements other methods for identifying AGN, such as
searching for broad emission-lines or hard X-rays, and has the
advantage of not being affected by obscuration.

\subsection{Radio Cores in LLAGN  -- the Grand Perspective}
\citeme{FalckeNagarWilson2000,FalckeNagarWilson2000b}
Assuming the cores detected in the VLA \& VLBA survey are produced by
randomly oriented, maximally efficient jets from supermassive black
holes (of order $10^8 M_\odot$) we can use Eq.~\ref{jetpower} to
calculate a minimum {\em total} jet power. For an average
monochromatic luminosity of $10^{27.5}$ erg sec$^{-1}$ Hz$^{-1}$ at 5
GHz the jets would have powers of order $Q_{\rm jet}\ga10^{42.5}$ erg
sec$^{-1}$. The way the model was constructed one has to consider this
a minimum energy estimate. Still, compared to quasars this is a rather low
value and supports the conclusion, based on their low UV and
emission-line luminosities, that the cores are powered by under-fed
black holes. On the other hand this jet power is well within the range
of the bolometric luminosity of typical low-luminosity AGN
($10^{41-43}$ erg sec$^{-1}$;~\citeNP{Ho1999}) and, compared to
radiation, jets should be a significant energy loss channel for the
accretion flow.

This similarity between jet power and (accretion) luminosity, of
course, appeals again to the jet-disk symbiosis picture discussed at the
beginning of this work. We can now take the radio cores detected in
our survey and put them on the correlation predicted in
\citeN{FalckeBiermann1996}. A problem one encounters is how to
estimate the accretion disk luminosity and accretion power. To make
the different AGN comparable we will use the narrow H$\alpha$ line
that is measurable in all AGN, and apply the same conversion factor
between H$\alpha$ and $L_{\rm disk}$ as used for quasars. For narrow
H$\alpha$ Eq.~\ref{oiii2uv} then reads

\begin{equation}
\lg (L_{\rm disk}/{\rm erg\;s}^{-1})=4.85+\lg (L_{\rm H\alpha,n}/{\rm erg\;s}^{-1}).
\end{equation}

\begin{figure}
\centerline{
\psfig{figure=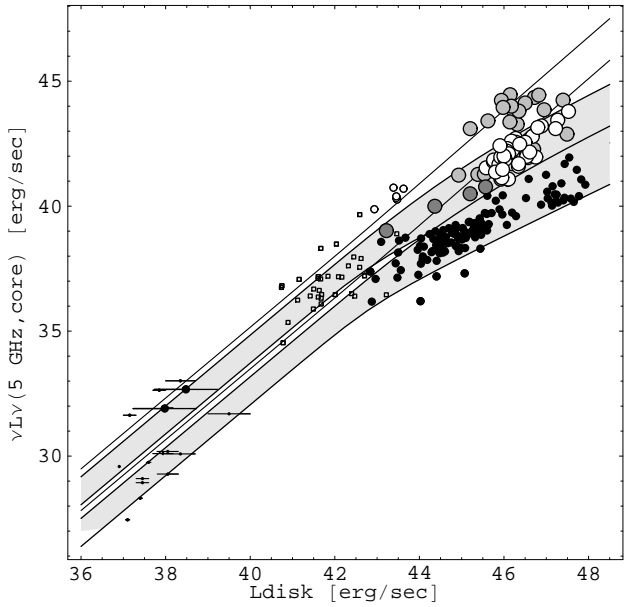,width=0.7\textwidth,bbllx=3.4cm,bblly=17cm,bburx=13.7cm,bbury=27cm,clip=}
}
\caption[]{\label{theplot-all}
Correlation between thermal emission from the accretion disk (with the
exception of X-ray binaries this is basically normalized to the
narrow H$\alpha$ emission) and the monochromatic luminosity of black
hole radio cores. Open circles: Radio-loud quasars; small open
circles: FR\,I radio galaxies; open gray circles: Blazars and
radio-intermediate quasars (dark grey); black dots: radio-quiet
quasars and Seyferts; small dots: X-ray binaries; small boxes:
detected sources from the ``48 LINERs'' sample. The latter apparently
confirm the basic prediction of Falcke \& Biermann (1996) and almost
close the gap between very low (on an absolute scale) accretion rate
objects and high accretion rate objects.  The shaded bands represent
the radio-loud and radio-quiet jet models as a function of accretion
as shown in Falcke \& Biermann (1996).}
\end{figure}

This estimate of course underestimates the accretion power or
accretion rate compared to quasars if LLAGN are largely dominated by
inefficient accretion, such as ADAFs.  Figure \ref{theplot-all} shows
the predicted Radio/$L_{\rm disk}$ correlation together with LLAGN
found in our survey. The galaxies almost fill the gap between quasars
and X-ray binaries on this absolute scale and roughly fall in the
predicted range. This illustrates that we have a continuation from
high-luminosity to low-luminosity AGN, {\em the latter being the
silent majority within the AGN family}.

\nocite{FalckeBiermann1996}

\begin{figure}
\centering
\noindent
\psfig{figure=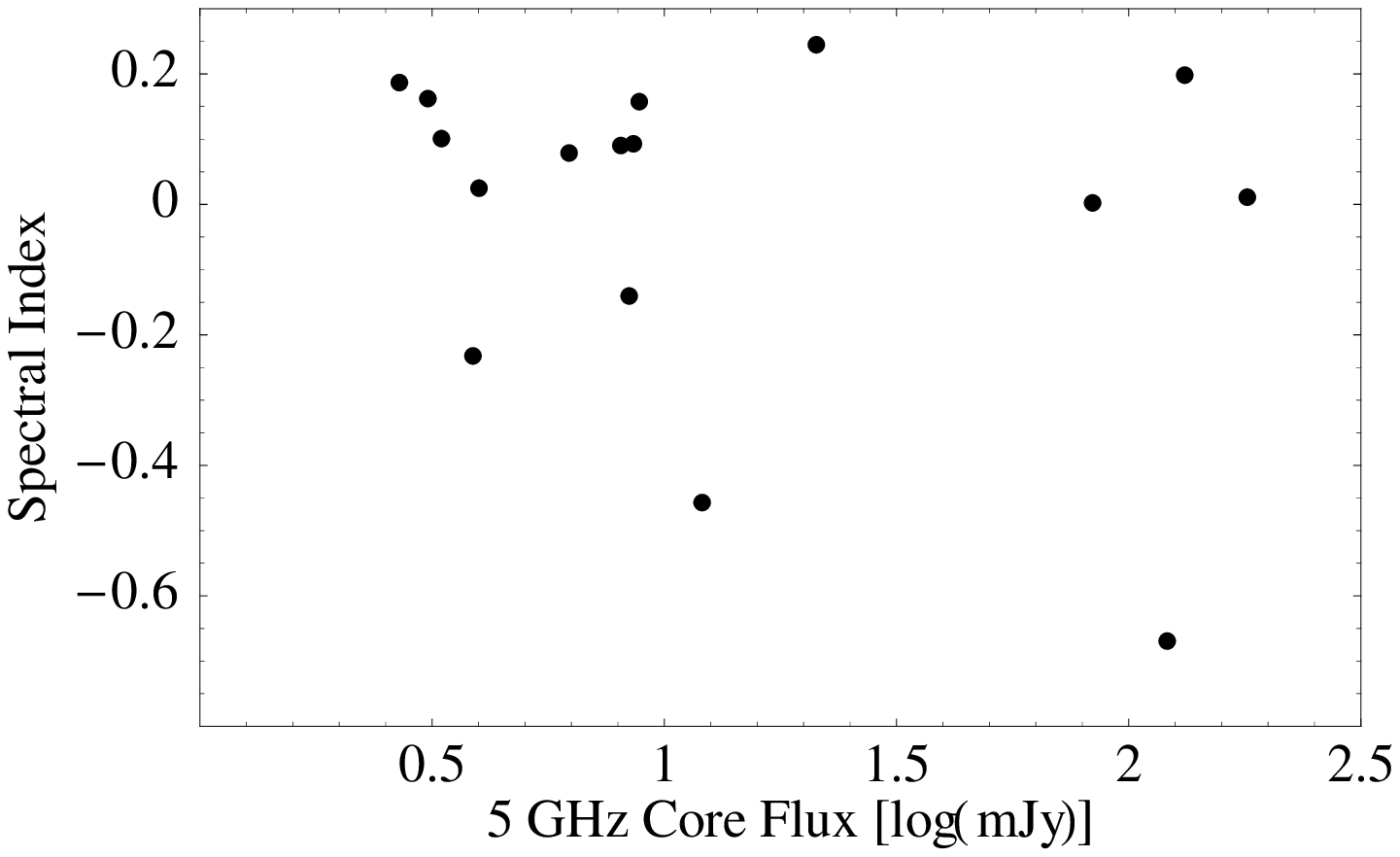,width=0.49\textwidth,bbllx=3.1cm,bblly=18.1cm,bburx=17.8cm,bbury=27cm,clip=}\psfig{figure=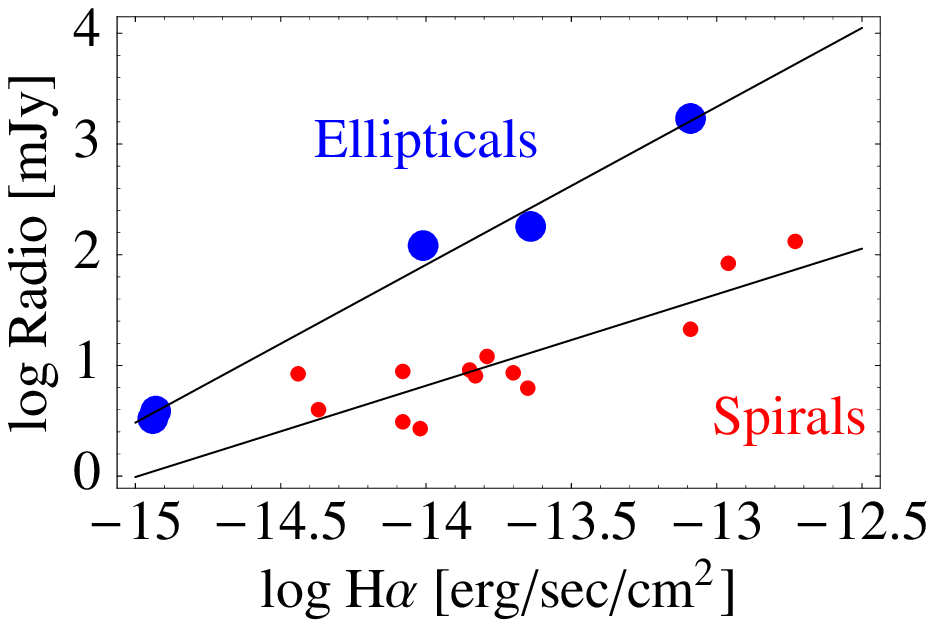,width=0.49\textwidth,bbllx=3.8cm,bblly=20.8cm,bburx=13.8cm,bbury=27cm,clip=}
\caption[]{Left: Spectral indices of LLAGN in our two samples with $S_{\rm 15 GHz}>3$
 mJy between 5 GHz (VLBI) and 15 GHz (VLA) as a function of radio core
flux at 5 GHz. Right: $S_{\rm 15 GHz}$ plotted versus narrow H$\alpha$
flux for the same sample; ellipticals and spirals are distinguished by
big and small dots respectively.}
\label{LLAGN-ha}
\end{figure}

We can investigate the optical/radio correlation in greater detail.
For the VLBI-sample (Nagar et al., in prep.), i.e.~the well-detected
cores above 3 mJy in both samples, for which we have basically
established that the radio emission is AGN-related, we can look at
correlations between radio, emission-line, and bulge
luminosities. Figure \ref{LLAGN-ha} (right panel) shows that there is
a trend for galaxies with higher H$\alpha$ emission to have more
luminous radio cores. Interestingly, elliptical and spiral host
galaxies are offset from each other.  The same effect can be seen in
Fig.~\ref{theplot-all} where one finds a string of sources connecting
to FR\,I radio galaxies and falling somewhat above the top line
predicted by the model. This are the large elliptical galaxies which
are probably faint versions of radio galaxies. Does this reflect a
radio-loud/radio-quiet dichotomy for LLAGN? 

In this respect it is noteworthy that both populations seem to lie
somewhat above the expected theoretical radio-optical
distribution. The radio cores in ellipticals are most likely the
continuation of FR\,I radio galaxies at low powers. For FR\,I radio
galaxies it is known that they seem to be underluminous in emission
lines
(e.g. \citeNP{FalckeGopal-KrishnaBiermann1995,ZirbelBaum1995}). If one
would artificially increase $L_{\rm disk}$ by a factor of about 30, the
radio cores in ellipticals and FR\,Is would nicely connect to those of
FR\,II radio galaxies and radio loud quasars. Applying the same factor
to the spiral galaxy cores would shift the remaining population into
the radio-quiet regime. Such a shift would be appropriate if, for a
given accretion rate or jet power, the radiative efficiency of the
disk would be reduced by such a factor (e.g. as in some ADAF models).

\begin{figure}
\centering
\noindent
\psfig{figure=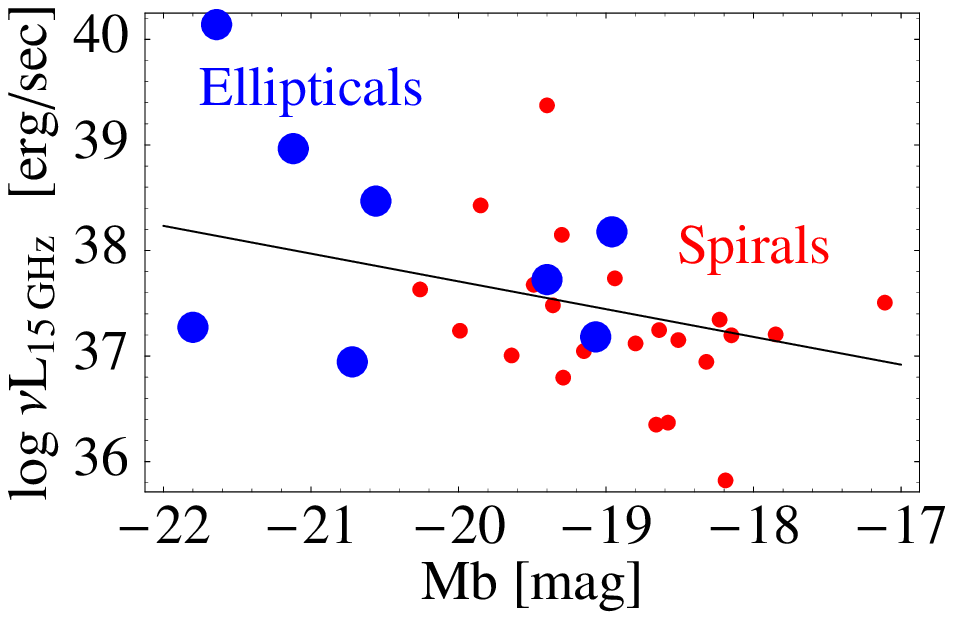,width=0.475\textwidth,bbllx=3.8cm,bblly=20.8cm,bburx=13.4cm,bbury=27cm,clip=}\psfig{figure=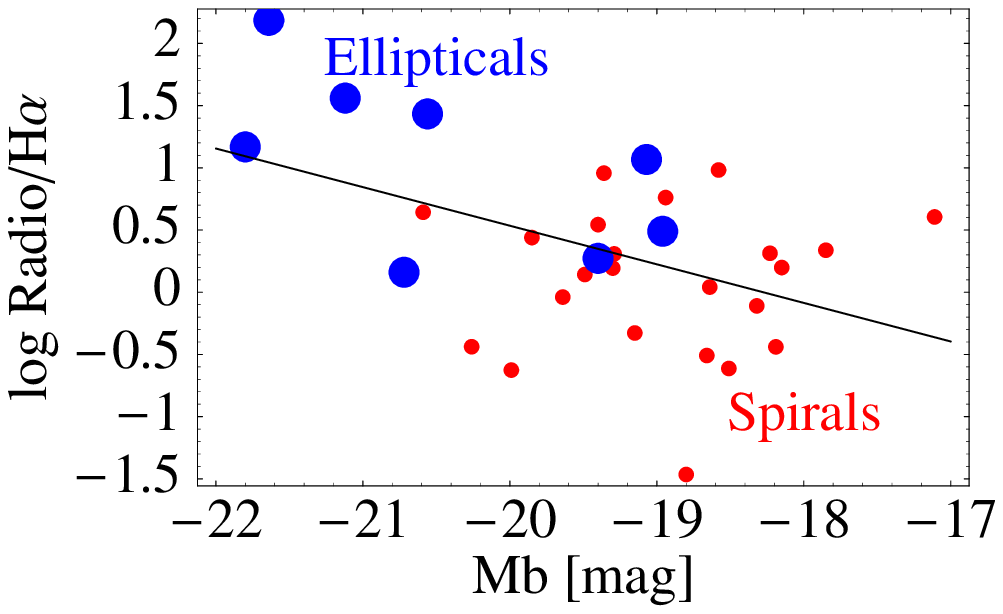,width=0.505\textwidth,bbllx=3.5cm,bblly=20.9cm,bburx=13.7cm,bbury=27cm,clip=}
\caption[]{Left: Radio luminosity ($\nu L_\nu$) at 15 GHz of LLAGN in our sample 
with $S_{\rm 15 GHz}>1.5$ mJy as a function of blue bulge
magnitude. Right: Ratio between 15 GHz radio core and narrow H$\alpha$
flux as a function of blue bulge magnitude in the same sample.
Ellipticals and spirals are distinguished by big and small dots
respectively.}
\label{LLAGN-Mb}
\end{figure}

There is, however, another important factor: the galaxy bulge
luminosity. We do see a weak trend for the radio luminosity to be
related to bulge luminosity; also the ratio between radio and
H$\alpha$ luminosity tends to increase with increasing bulge
luminosity. Hence, galaxies apparently become more efficient in
producing radio emission relative to H$\alpha$ in bigger bulges. This
also holds if we look at the entire VLA detected sample
(Fig.~\ref{LLAGN-Mb}). Whether this is due to increasing obscuration,
effects intrinsic to the AGN, or a selection effect is unclear.  In
any case, we may be comparing pumpkins with apples. Since ellipticals
and spirals in our sample are nicely separated between the top and
bottom end of the bulge luminosity distribution, an apparent dichotomy
in Fig.~\ref{LLAGN-ha} is a natural consequence of this trend.

If the bulge luminosity is proportional to the central black hole mass
\cite{KormendyRichstone1995,FerrareseMerritt2000,GebhardtBenderBower2000},
the ellipticals in our sample are more likely to have larger black
hole masses. This means they have a larger 'headroom' with respect to
their Eddington limit and are more likely to have larger accretion
rates. Moreover, if an accretion disk becomes radiatively less
efficient the further away it is from the Eddington limit, we could
reproduce the scaling of the Radio/H$\alpha$-ratio. Another
explanation for the latter could be that the larger bulges of these
ellipticals contain more obscuring material towards our line of sight
and hence optical emission is more strongly suppressed.

To summarize: we find that at least 40\% of optically selected LLAGN
with Seyfert and LINER spectra have compact radio cores.  VLBI
observations show that these cores are similar to radio jets in more
luminous AGN with high brightness temperatures, jet-like structures,
and flat radio spectra. The radio emission seems to be related to the
luminosity of the emission-line gas and hence both are probably
powered by genuine AGN operating at low powers. We found no evidence
for high frequency components with highly inverted spectra as predicted
in ADAF models. Hence, for these models one should probably not
include radio fluxes in broad-band spectral fits. 

We also find only a weak correlation between radio and bulge
luminosity, which could imply a scaling of radio emission with black
hole masses.  Such an effect was claimed earlier and interpreted
within the ADAF models as a possibility to measure black hole masses
with the radio data \cite{FranceschiniVercelloneFabian1998}. However,
this result was based on a much smaller, ill-selected sample and gave
a much steeper dependence which we cannot confirm. As we have seen
here, the radio-H$\alpha$ correlation is a much stronger effect. If
the H$\alpha$ is a tracer of the accretion disk luminosity, this most
likely means that the radio power is much more dependent on the
accretion rate than on the black hole mass. One will therefore have to
continue to measure black hole masses in the traditional way, mainly
through spectroscopy.

\section{Summary}

Radio jets are the ``smoking guns'' of active galactic nuclei
(AGN). They are also the site of many high-energy processes, including
X-ray and $\gamma$-ray emission as well as high-energy particle
acceleration. Early on it was the radio emission from these jets which
drew people's attention to quasars and led to their discovery. Today
we know that radio jets are relativistic, magnetized plasma flows
ejected at velocities close to the speed of light most likely from the
immediate vicinity of black holes. At the smallest scales, the radio
emission from these jets appears as very compact radio cores with a
flat and variable radio spectrum. With Very Long Baseline
Inter\-ferometry this emission can often be resolved into a core-jet
structure sometimes showing apparent superluminal motion caused by the
relativistic ejection velocity.

However, similar radio cores have also been found in quite a number of
other sources, including the center of the Milky Way (Sgr A*), X-ray
binaries (stellar mass black holes and neutron stars; see also
\citeNP{MarkoffFalckeFender2000,FenderHendry2000}), and some nearby
galaxies. Despite being interesting in their own right, none of these
sources can compare in their power output and violence with
quasars. On the other hand it is thought that in all these cases black
holes (and in a few cases also neutron stars) are involved as well,
possibly accreting at much lower levels -- in absolute terms -- than
quasars.

Here we have discussed how these low-luminosity radio cores are
related to their much more powerful siblings in quasars, radio
galaxies, and Blazars. It is suggested that all radio cores can be
described in a very similar fashion as jets coupled to an accretion
disk (standard-$\alpha$ or ADAF), with jet and disk being symbiotic
features in an accretion scenario of black holes. At some level every
black hole will accrete material but only a minority of sources reach
the enormous power levels of quasars. The ``silent majority'' of black
holes therefore operates at much lower levels producing much fainter
radio cores.

Based on this idea of a ``jet-disk symbiosis'' a general emission
model for radio cores can be derived which describes the radio
emission of a collimated relativistic plasma flow after leaving its
collimation region -- the nozzle. This part of the jet is most likely
dominated entirely by free expansion and is relatively independent of
environmental influences. This leads to a relatively homogeneous
appearance of compact radio cores in different sources with the jet
power or the accretion rate being the main parameter. Such a model can
explain sizes, fluxes, and spectra of many radio cores in weakly
active black holes, such as Sgr A* in the Galactic Center or other
Low-Luminosity AGN.

To conclude, one can say that the production of a relativistic jets
seems to be an inevitable consequence of accreting black holes and is
possible even at very low accretion rates. Therefore, compact radio
cores are an ideal tracer of black holes in the near and distant
universe. With the increasingly higher resolution and sensitivity of
radio interferometers, they allow an intimate look at how black holes
work at various scales and in different contexts. 

\subsection*{Danksagungen}

Ich m\"ochte mich ganz herzlich bei dem Vorstand der Astronomischen
Gesell\-schaft, insbesondere bei dem Vorsitzenden Herrn
Prof.~Sedlmayr, f\"ur die Verleihung des Ludwig-Biermann-Preises
und die Einladung nach Bremen bedanken.\ Ebenso m\"ochte ich an
dieser Stelle den Herren Dr.\ Anton Zensus und Prof. Dr. Peter L.
Biermann f\"ur ihre Unterst\"utzung und F\"orderung in den letzten
Jahren danken.

Herzlich danke ich auch meinen Kollegen, die zu den hier
vorgestellten Arbeiten wesentlich beigetragen haben:\ Dr.~Geoffrey
Bower, Dr.~Sera Markoff, Dr.~Eric Agol, Prof.~Fulvio Melia,
Dr.~Neil Nagar und Prof.~Andrew Wilson.\ Diese Arbeit ist eine
stark verk\"urzte und \"uberarbeitete Fassung meiner
gleich\-namigen Habilitationsschrift.\ Dort ist auch eine
umfassende Liste zugrundeliegender Originalarbeiten (soweit
erschienen) sowie eine deutschsprachige Zusammenfassung zu finden.

\vspace{0.7cm}
\noindent
\bibliographystyle{apj}
\bibliography{apjmnemonic,../../Review/review}

\end{document}